\documentclass{aa}  
\usepackage{booktabs}
\usepackage{comment}
\usepackage{xcolor}
\usepackage{xspace}
\usepackage[normalem]{ulem}

\usepackage{natbib}
\bibpunct{(}{)}{;}{a}{}{,} % to follow the A&A style

\newcommand*{\cf}{cf.\@\xspace}
\newcommand*{\ie}{i.e.\@\xspace}
\newcommand*{\eg}{e.g.\@\xspace}

\newcommand{\mghk}{\ion{Mg}{II} h\&k\xspace}
\newcommand{\mgh}{\ion{Mg}{II} h\xspace}
\newcommand{\mgk}{\ion{Mg}{II} k\xspace}

\newcommand{\konev}{$\mathrm{k_{1v}}$\xspace}
\newcommand{\koner}{$\mathrm{k_{1r}}$\xspace}
\newcommand{\kone}{$\mathrm{k_{1}}$\xspace}
\newcommand{\ktwor}{$\mathrm{k_{2r}}$\xspace}
\newcommand{\ktwov}{$\mathrm{k_{2v}}$\xspace}
\newcommand{\ktwo}{$\mathrm{k_{2}}$\xspace}
\newcommand{\kthree}{$\mathrm{k_{3}}$\xspace}

\newcommand{\htwo}{$\mathrm{h_{2}}$\xspace}

\newcommand{\cahk}{\ion{Ca}{II} H\&K\xspace}

\newcommand{\halpha}{\mbox{H\hspace{0.1ex}$\alpha$}\xspace}

\newcommand{\muram}{MURaM-ChE\xspace}
\newcommand{\bifrost}{Bifrost\xspace}
\newcommand{\iris}{IRIS\xspace}
\newcommand{\hmi}{HMI\xspace}

\newcommand{\snapfourNineNine}{\texttt{muram\_en\_499000\_379s}\xspace}

\newcommand{\km}{\ensuremath{\, \mathrm{km}}\xspace}
\newcommand{\kG}{\ensuremath{\, \mathrm{kG}}\xspace}
\newcommand{\G}{\ensuremath{\, \mathrm{G}}\xspace}
\newcommand{\tauUnity}{\ensuremath{\tau_{500}=1}\xspace}
\newcommand{\kms}{\ensuremath{\, \mathrm{km\,s^{-1}} }\xspace}
\newcommand{\Mm}{\ensuremath{\, \mathrm{Mm}}\xspace}
\newcommand{\nm}{\ensuremath{\, \mathrm{nm}}\xspace}

\usepackage{graphicx}
\usepackage{txfonts}
\begin{document}

\title{\ion{Mg}{II} h\&k spectra of an enhanced network region simulated with the \muram code}
\subtitle{Results using 1.5D synthesis}
\author{ P. Ondratschek\inst{1}, D. Przybylski\inst{1}, H.N. Smitha\inst{1}, R. Cameron\inst{1}, S.K. Solanki\inst{1}, J. Leenaarts\inst{2}}
\authorrunning{Ondratschek et al.}
\titlerunning{\mghk spectra, 1.5D RT results}

\institute{Max-Planck Institute for Solar System Research, 37077 G\"{o}ttingen, Germany
            \mail{ondratschek@mps.mpg.de}
            \and Institute for Solar Physics, Department of Astronomy, Stockholm University, AlbaNova University Centre, 106 91 Stockholm,
Sweden
          }

\abstract
{% Context
The \mghk lines  are key diagnostics of the solar chromosphere. They are sensitive to the temperature, density, and non-thermal velocities in the chromosphere. The average \mghk line profiles arising from previous 3D chromospheric simulations are too narrow.
}
{% Aims
We study the formation and properties of the \mghk lines in a model atmosphere. We also compare the average spectrum, peak intensity, and peak separation of \mgk with a representative observation taken by the Interface Region Imaging Spectrograph (IRIS).
}
{%Methods:
We use a model based on the recently developed non-equilibrium version of the radiative magneto-hydrodynamics code MURaM, the MURaM Chromospheric Extension (\muram), in combination with forward modeling using the radiative transfer code RH1.5D to obtain synthetic spectra. Our model resembles an enhanced network region created by using an evolved MURaM quiet sun simulation and adding a similar imposed large-scale bipolar magnetic field as in the public Bifrost snapshot of a bipolar magnetic feature. }
{%Results:
The line width and the peak separation of the spatially averaged spectrum of the \mghk lines from the \muram simulation are close to a representative observation from the quiet sun which also includes network fields. However, we find the synthesized line width to be still slightly narrower than in the observation.
We find that velocities in the chromosphere play a dominant role in the broadening of the spectral lines. While the average synthetic spectrum also shows a good match with the observations in the pseudo continuum between the two emission lines, the peak intensities are higher in the modeled spectrum. This discrepancy may partly be due to the larger magnetic flux density in the simulation than in the considered observations but also due to the 1.5D radiative transfer approximation. }
{%Conclusions:
Our findings show that strong maximum velocity differences or turbulent velocities
in the chromosphere and lower atmosphere are necessary to reproduce the observed line widths of chromospheric spectral lines.}


   \maketitle

%-----------------------------
\section{Introduction}                                                        \label{sec:introduction}
%--------------------
    The solar chromosphere is highly dynamic and complex. Even though an exact definition of this part of the solar atmosphere is not straightforward, it is spatially located above the photosphere and below the transition region which connects the chromosphere to the corona. It can also be described as the region above the solar surface where hydrogen is partially ionized but still mostly neutral. In order to study the energy transfer in the solar atmosphere a detailed knowledge of the structure and dynamics of the chromosphere is essential. Much about the chromosphere is still not well understood. For example, the role of shock waves in structuring and heating the chromosphere versus the role of magnetic fields is not fully resolved. Interpretations of observed chromospheric phenomena, for example through the \mghk lines, require a comprehensive understanding of the underlying atmosphere.

Observationally, chromospheric spectral lines are the key to investigate the chromosphere in detail as they contain signatures of the plasma conditions in their formation regions. Strong chromospheric lines can be observed from the ultraviolet (UV) to the infrared. These include well known lines such as \mghk, \cahk, \halpha, and the \ion{Ca}{II} infrared triplet. In this work we focus on the \mghk lines which are centered at 2796 \AA{} (\mgk) and 2831 \AA{} (\mgh), and are thus inaccessible from ground-based telescopes as the UV part of the spectrum is blocked by the Earth's atmosphere. 
 The first high resolution images in the core of the \mgk line \citep{riethmueller_2013ApJ...776L..13R,danilovic_2014ApJ...784...20D} were taken by the SuFI instrument \citep{gandorfer_2011SoPh..268...35G} onboard the second flight of the Sunrise balloon-borne observatory \citep{solanki_sunrise_2010ApJ...723L.127S,solanki_sunrise_first_results_2012ASPC..455..143S,barthol_sunrise_2011SoPh..268....1B}. 
Currently, the main source of long-term \mghk observations is the Interface Region Imaging Spectrometer \citep[\iris,][]{Pontieu2014_iris} satellite launched in 2013.

In the present paper, our aim is to compute and analyze the \mghk lines in a new model of the solar chromosphere and to compare the resulting profiles with \iris observations.
Despite using advanced techniques to model the chromosphere, and to compute the emergent intensity, discrepancies are still found between observed spectra and those arising from MHD simulations. Computed \mghk spectra show on average too narrow line widths, and too weak peak intensities \citep{carlsson_chromosphere_review2019}. A number of possible explanations have been proposed. \cite{Leenaarts_2013_mgii_hk_bifrost_diagnostics} posed that these effects might be coming from too weak velocity fields  and from an underestimated mid-chromospheric temperature in the models. \cite{Carlsson2016a_public_bifrost_snapshot} argued that the magnitude of "non-thermal velocity" in the atmosphere, which contributes to the broadening of spectral profiles, can be increased by using a higher resolution simulation domain. \citet{Carlsson_2015_mg_plage}, \citet{carlsson_chromosphere_review2019}, and \citet{hansteen_numerical_mghk_2023ApJ...944..131H} argued that besides macroscopic velocities, the so-called opacity broadening \citep{Rathore_2015_opacity_broadening} affects the width of the \mghk lines.
\\ \indent
Propagating shocks, decreasing density, and magnetic fields lead to a highly structured chromosphere. This is reflected in the formation of \mghk. Two dominant effects are the non local-thermodynamic-equilibrium (NLTE) formation of the spectral lines and partial frequency redistribution (PRD). NLTE conditions lead to deviations of the line source function from the Planck function. The second effect is a result of scattering which becomes important in the upper chromosphere \citep[for \mgk \eg, see][]{milkey_mihalas_mg_prd_1974ApJ...192..769M,Leenaarts_2013a}. PRD effects in the formation of \mghk have also been studied in the context of scattering polarization \citep[see \eg][]{auer_scattering_polarization_1980A&A....88..302A,beluzzi_mghk_scattering_PRD_2012ApJ...750L..11B}.
\\ \indent
In order to study line formation in the chromosphere a model atmosphere is required, which provides temperatures, hydrogen populations, and electron densities accounting for NLTE effects. Early studies on the formation of \mghk in solar-like atmospheres were based on one-dimensional (1D) models. The semi-empirical models of \citet[][VAL models]{Vernazza_avrett_loeser_VAL_A-F_1981ApJS...45..635V} or \citet[][FAL models]{fontenla_FALC_1993ApJ...406..319F} represent spatial and temporal average conditions of the solar atmosphere. More sophisticated hydrodynamics codes like RADYN \citep{carlsson_1992_RMHD,radyn_2002ApJ...572..626C} are used to study dynamic atmospheres including propagation of waves and shocks and a treatment of radiation transfer for determining the non-equilibrium populations and NLTE losses. Self-consistent models need 3D geometry to accurately model the convection and dynamics of the atmosphere, in addition to 3D radiation transfer \citep{leenaarts_2020_radiation}, because of the scattering in the chromosphere. \bifrost \citep{Gudiksen_2011_Bifrost} is a radiation magnetohydrodynamics (rMHD) code that fulfills these and the above-mentioned requirements. \\ \indent
In this work, we use the recently developed chromospheric extension of the MURaM code \citep{Przybylski2022a_muram_chromospheric_extension}, which we will refer to in the following as \muram.\footnote{Max Planck Institute for Solar System Research/University of
Chicago Radiation Magneto-hydrodynamics with the chromospheric extension.}
The chromospheric extension includes a non-equilibrium (NE) treatment for hydrogen ionization and approximations to NLTE radiative line losses.
The simulated atmosphere is a reproduction of the Bifrost enhanced network simulation \citep[][hereafter the public \bifrost snapshot]{Carlsson2016a_public_bifrost_snapshot}, with a similar large scale field. We use the atmosphere model to compute synthetic spectra of the \mghk lines in the "plane-parallel" approximation where each vertical column of the 3D atmosphere is treated as an individual 1D atmosphere for the radiative transfer (RT) computation (also called 1.5D approximation). Horizontal components of the velocity field are neglected in this approximation.
\\ \indent
In Sect. \ref{sec:methods} we describe the numerical tools, including the MHD simulation, spectral synthesis, and the observations. In Sect. \ref{sec:results} we present the results from the forward modeled spectra and compare them with observations and the public \bifrost snapshot. A summary and conclusions are presented in Sect. \ref{sec:discussion}.

%=============================
\section{Methods}                                                                    \label{sec:methods}
%===============================

In this Sect. we describe the methods on which our analysis is based. We briefly describe the \muram code and the setup of our enhanced network model. We then describe how the RT computations were performed. After this we  introduce the observations and the public \bifrost snapshot used for comparison. Followed by this we describe the procedure to spatially and spectrally degrade the synthetic spectra, in order to better match the conditions imposed on the observations by the instrumentation, and finally present a description of how the different spectral features were identified.

%----------------------------
\subsection{Enhanced Network Model}                                   \label{sec:methods_enhanced_network_model}
%-----------------------------

We use results of an rMHD simulation performed with the chromospheric extension of the MURaM code %
\citep{Przybylski2022a_muram_chromospheric_extension}.
The code is based on the LTE radiative magnetoconvection code by 
\citet{Voegler2015},
extended to include optically thin losses and point-implicit heat conduction by
\cite{rempel_coronal_extension_2017ApJ...834...10R} in order to treat the physics of the corona. 

The simulation setup is similar to that described in \citet{Przybylski2022a_muram_chromospheric_extension}, and includes a 4-band multigroup RT scheme \citep{Nordlund_1982_theOG,voegler_2004_approximations} extended to include scattering effects \citep{skartlien_2000_multigroup, hayek_2010_radiative}. Additional radiative losses from strong chromospheric lines and optically thin losses in the corona are included as described in \cite{carlsson_2012_approximations}. Furthermore, we include 3D extreme-ultra-violet (EUV) back-heating of the chromosphere similar to \cite{carlsson_2012_approximations}.
In the convection zone we use a non-ideal Equation-of-State (EoS) generated with the free-EoS package \citep{Irwin_2012_freeeos}, joined to a time-dependent, non-equilibrium treatment of hydrogen in and above the photosphere \citep{sollum_thesis_1999, leenaarts_2006_timedependant, leenaarts_2007_nonequilibrium}. We use abundances from \citet{asplund_abundances_2009}. The time-dependent EoS includes a non-equilibrium treatment of molecular $H_2$, and  $H_2^{+}$ and $H^{-}$ in chemical equilibrium. In the EoS all non-hydrogen elements are treated in LTE. 
The diffusion scheme follows \cite{rempel_2014_numerical, rempel_coronal_extension_2017ApJ...834...10R}, which is based on a slope-limited scheme. Numerical diffusion is included based on monotonicity constraints, which are disabled in sufficiently smooth regions. This allows the simulation to run smoothly with minimal resistivity and viscosity. In addition, we used the "Boris" correction \citep{Boris_1970_BC} to limit the Alfv\'en speed, which enables simulations at high numerical resolution to be run with a reasonable computational cost by relaxing the Alfv\'en speed contribution to the Courant-Friedrichs-Lewy (CFL) criterion \citep{courant_uber_1928_CFL}.

\begin{figure}
   \centering
   \includegraphics[width=\hsize,clip]{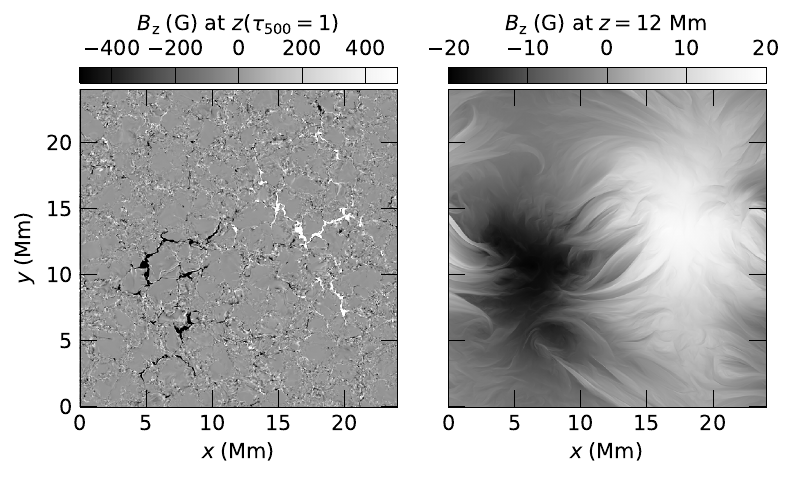}
      \caption{Left panel: Vertical magnetic field map of the \muram simulation at the $\tau_{500}=1$ layer.  The gray scale is saturated at $|B_z| = 500 \, \mathrm{G}$ to pronounce some of the weaker internetwork fields. Right panel: Vertical magnetic field at a constant height of $12 \Mm$ with a gray scale saturated at $|B_z| = 20 \, \mathrm{G}$.}
         \label{fig:bzphotosphere}
\end{figure}

The setup of the bipolar structure for an enhanced network environment is close to the public \bifrost snapshot. Due to differences between the EoSs in the \muram and \bifrost codes, a direct restart of the \bifrost public snapshot did not result in a sun-like star. To build the model used in this paper we use an existing small-scale dynamo (SSD) model (Przybylski et al., in preparation). The SSD model extends $12 \Mm \times 12 \Mm \times 18 \Mm$ with a resolution of $23.46 \km$ horizontally and $20 \, \km$ vertically. The atmosphere ranges from $-6.6 \Mm$ below to $11.4 \Mm$ above the photosphere.
We tiled this model 2x2, by copying the simulation box three times, to produce a box that extends $24\Mm \times 24\Mm$ horizontally with 1024x1024 grid points.  The domain was extended 4 Mm further in the vertical direction, to give a total extent from ~-7 \Mm to 17 Mm with 1200 grid points (with the solar surface lying at a height of 0 \Mm). The resolution is therefore the same as in the SSD model. The \muram code uses an equidistant vertical axis, that is the grid is not variable as in other codes (\eg \bifrost).
This model is run for 8 hours, to break the periodicity, which comes from tiling the smaller SSD model to the larger simulation domain, and to allow the simulation to relax from transients due to the extension. The horizontal boundary conditions are periodic. The upper boundary is open to outflows but closed to inflows. The magnetic field at the top boundary is treated as a potential field. In the lower boundary we use an open symmetric-field condition as described in \cite{rempel_2014_numerical}. We added a bipolar magnetic field to the simulation, to give a large scale field similar to the initial condition of the public Bifrost snapshot on top of the existing SSD model. We then ran the model for additional $1.5$ hours. After that we turned on the non-equilibrium treatment of hydrogen and ran the computations for another 10 minutes in order to allow the populations to settle into a new equilibrium. We then saved a 10 minute time series at $\approx 6~\mathrm{s}$ cadence, that we use for our analysis. Additionally, we saved 8 2D slices at fixed $x$ and $y$ positions through the box, but with a higher temporal cadence of 1.2 seconds.
In Fig. \ref{fig:bzphotosphere} we show the vertical component of the magnetic field at the $\tau_{500}  = 1$ plane and at a constant height of $12 \Mm$ for the snapshot discussed in Sect. \ref{sec:results}. The absolute value of the vertical component of the magnetic field at the photosphere can reach values of up to $|B_{z}(\tauUnity)| \approx 2 \kG$, but we clipped the gray scale to values of $\pm 500 \G$ to also highlight the smaller structures. The spatial unsigned mean of the vertical component of the magnetic field is $\langle|B_z (\tauUnity)| \rangle = 64 \G$.

In observations, as for example taken by the Helioseismic and Magnetic Imager \citep[HMI,][]{scherrer2012_hmi,schou_design_2012}, the Stokes signal of the \ion{Fe}{i} $\lambda 6173\, \AA$ line is used to measure the line-of-sight (LOS) magnetic field. The wings of the \ion{Fe}{i} $\lambda 6173\, \AA$ line, where the Stokes-V signal is strongest, are expected to form approximately at $\tau_{500} = 0.1$ \citep[\eg,][]{fleck_hmi_2011SoPh..271...27F, Quintero_Noda_2021A&A...652A.161Q}. We therefore degraded the magnetogram of the simulation at $\tau_{500} = 0.1$ to a resolution of $1"\approx 760 \km$, by convolving the data by a Gaussian kernel, to make it comparable in resolution to an HMI magnetogram  taken during the IRIS observation we compare our synthetic spectra with (see Sect. \ref{sec:methods_observations}). The degradation in spatial resolution led to $\langle|\Tilde{B}_z (\tau_{500}=0.1)| \rangle = 20.7 \G$, where $\Tilde{B_z}$ refers to the  field strength after spatial degradation. For comparison, the value computed at the original resolution is $\langle|B_z (\tau_{500}=0.1)| \rangle = 43.5 \G$.

 The left panel of Fig. \ref{fig:bzphotosphere} shows the magnetogram at the \tauUnity plane where small magnetic structures arising through the SSD action can be seen. The large scale bipolar structure appears along the horizontal diagonal, that is from the lower left to the upper right corner in Fig. \ref{fig:bzphotosphere}a,b, of the simulation domain and dominates at a constant height of $12 \Mm$ (right panel).

%---------------------------------

We want to emphasize that even though the \muram and \bifrost models appear to be similar in terms of the numerical treatment and the initial conditions, there are significant differences in the resolution, simulation depth, top boundary condition, EoS, and diffusion scheme. Therefore, a direct comparison is not straightforward and is not the subject of this work.

%---------------------------------------
\subsection{Synthetic spectra}                                                    \label{sec:methods_synthesis}
%-----------------------------------

We perform detailed RT calculations to model the \mghk emerging intensity for comparison with observations. We used the RT code RH1.5D \citep{pereira_2015_rh1p5d} which is based on the original RH code \citep{Uitenbroek_2001}. The code solves the radiative transfer problem using the multilevel accelerated lambda iteration (MALI) method as described in \cite{Rybicki_Hummer_ALI_1991, uitenbroek_2002_mali_1992A&A...262..209R}. The \mghk lines are computed in NLTE and the angle-dependent PRD effects are taken into account via the hybrid approximation described in \cite{Leenaarts_2012b_hybrid_prd_approximation}. We use the 10+1 level \ion{Mg}{II} atom described in \cite{Leenaarts_2013a}. Other elements such  as H, C, O, Si, Al, Ca, Fe, He, Ni, Na, and S were included in the background by treating them in LTE. Similar to \cite{Pereira2013_iris_diagnostics_iii} we additionally included the strongest 15 \%  of the lines from the Kurucz line list  \citep{kurucz1995} in the IRIS NUV window. These blending lines are mostly located in the inner wings and the pseudo-continuum between the \mghk lines, and a few of these blends lie close to the cores of the Mg lines. However, they are not strong enough to have an impact, neither on the intensities nor on the shapes of the \mghk line cores. 
The abundances were taken from \cite{asplund_abundances_2009}.
We used a maximum  relative change in the atomic level population of $|\Delta n/n| =  10^{-3}$ as the stopping criteria for the iterative solution. In the RT computation the impact of magnetic fields on the emergent intensity through the Zeeman effect was neglected. The 1.5D RT computation does not take horizontal velocity components into account. This means there is no potential reduction in intensity contrast as for example in \citet{bestard_2021ApJ...909..183J} in the case of \ion{Ca}{i} $\lambda 4227\AA$. The source function was interpolated using a linear scheme. This allows for a better comparison with the Bifrost spectra (see IRIS Technical Note, 35\footnote{IRIS technical note (ITN) 35: \url{https://www.lmsal.com/iris_science/doc?cmd=dcur&proj_num=IS0217&file_type=pdf}}). In addition, we did not find a significant difference between the linear  and a cubic Bezier interpolation scheme.

In order to reduce the computational costs, we cropped the vertical extent of each column in the atmosphere at a height along the LOS where a temperature of $T_{\mathrm{cut}}=50 \, \mathrm{kK}$ is reached. The height where the atmosphere is cut is determined by going from the top of the atmosphere downwards until the value of the temperature is lower than $T_\mathrm{cut}$. Furthermore, for three of the four snapshots considered here only every second column in both the $x$ and $y$ directions in the atmosphere was used for spectral computations. To justify the latter approximation we computed snapshot \snapfourNineNine\footnote{We use the following name convention for snapshots: muram\_en\_iterationNumber\_time  where the time is measured in sec from the beginning of the 10 min series that we used for our analysis.} in full resolution and compared the average spectrum with the one with only $25 \%$ of the pixels, finding no significant difference in the \mghk window and in the distributions of peak intensities and peak separations.

%-------------------------------------------------------------------------------------------------------
\subsection{Observations}                                              \label{sec:methods_observations}
%-------------------------------------------------------------------------------------------------------
We use data from the IRIS satellite. 
The NUV window of \iris covers a wavelength range from 2783–2834 \AA{} with a spectral resolution of $\approx 6 \, \mathrm{pm}$ and a spatial resolution of $0".4 \approx 240 \, \mathrm{km}$. For estimations of the magnetic field in the photosphere we use data from HMI on board the Solar Dynamics Observatory (SDO). HMI magnetograms provide the LOS magnetic field strength in the lower solar atmosphere estimated using Milne-Eddington inversions of the \ion{Fe}{i} $\lambda 6173.3 \,\AA$ spectral line profiles \citep{borrero_hmi_2011SoPh..273..267B}.

We use an \iris raster observation from a quiet region that includes network fields, in order to compare with the profiles from the \muram model. The observation was taken on 2014-06-07 and has a field of view of $139" \times 182"$ (or approximately $105 \Mm \times 138 \Mm$) with a cadence of $15$ s. The observation was taken close to disk center at $\mu = 0.96$, where $\mu=\cos(\theta)$ with $\theta$ being the heliocentric angle. While our synthetic spectra are computed at $\mu=1.00$ we do not expect this slightly inclined viewing angle to have a strong impact on the observed intensity \citep[see, \eg,][]{avrett_2013ApJ...779..155A,sukhorukov2017}. The field-of-view (FOV) is larger than the simulation domain of the model. In Appendix \ref{sec:app-additional-observations} we consider three smaller FOVs, that are similar in size to our model, together with two additional observations, demonstrating that the chosen dataset is representative enough for the purpose of this work.

Therefore, we show in Appendix \ref{sec:app-additional-observations} that the chosen dataset is representative for the purpose of this work. The data were available in a rebinned format such that the final spectral and spatial resolutions are half of the original \iris specifications. 

We aligned the HMI magnetogram, recorded roughly half way through the \iris observation, with the raster by performing the following steps. First, we projected the HMI magnetogram onto the \iris FOV. For this purpose we used the "reproject\_interp" function which is part of the Astropy Project \citep{astropy_2022ApJ...935..167A}. Then we corrected the HMI image and the \iris raster scan for the effect of solar rotation which we assumed to be only noticeable in the east-west direction, as the observations are close to disk center.

The averaged unsigned LOS magnetic field over the whole FOV, calculated from the HMI magnetogram and corrected for the viewing angle $\mu$, is $\langle|B_{\mathrm{LOS}} / \mu| \rangle = 12.12 \G$ which is smaller than the degraded value at $\tau_{500}=0.1$ of $20.7 \G$ from the simulation. However, recent studies have shown that the magnetic flux measured by HMI could be underestimated by not accounting for NLTE effects in the \ion{Fe}{i} $\lambda$6173.3 \AA\ line \citep{smitha_nlte_iron_2023A&A...669A.144S} or because magnetograms can miss a significant fraction of the magnetic flux even in unipolar magnetic field regions \citep{sinjan2024magnetogramsunderestimateunipolarmagnetic,milic_2024A&A...683A.134M}.
In Appendix \ref{sec:app-additional-observations} (Table \ref{app:tab-observations}) we show the mean unsigned LOS magnetic field strengths for smaller regions of interest (ROI) that are comparable in size to our simulation. Their values in the selected ROIs vary from $6.8 \G$ (quiet sun) to $33.2 \G$ (network).

%-------------------------------------------------------------------------------------------------------
\subsection{The public bifrost snapshot}                \label{sec:methods_bifrost_model}
%-------------------------------------------------------------------------------------------------------

The \bifrost model from \cite{Carlsson2016a_public_bifrost_snapshot} resembles an enhanced network region. It has a horizontal extent of $24 \Mm \times 24 \Mm$ and a vertical extension from $-2.4 \Mm$  below and $14.4 \Mm$ above the $\tau_{500} = 1$ surface. The horizontal spatial resolution of this simulation is $48 \, \mathrm{km}$ and the vertical resolution varies between $19 \km$ in the photosphere and chromosphere, and up to $100 \km$ at the top boundary. A bipolar magnetic field structure with 8 Mm separation between the opposite poles was added to the bottom of the simulation domain. At $t=3850 \, \mathrm{s}$ the average unsigned vertical magnetic field strength at the photosphere is 48 G. The public release\footnote{\url{http://sdc.uio.no/search/simulations?sim=en024048_hion}} of the enhanced network simulation also includes synthesized spectra of the \mghk lines which we use for comparison. These spectra were also computed with the plane-parallel 1.5D approximation and do not include a 3D RT treatment for the line centers, in contrast to the profiles described in \cite{Leenaarts_2013_mgii_hk_bifrost_diagnostics}.

%-------------------------------------------------------------------------------------------------------
\subsection{Spatial and spectral degradation}                           \label{sec:methods_degradation}
%-------------------------------------------------------------------------------------------------------
We degraded the spectra from both the \muram model and the public \bifrost snapshot to the spatial and spectral resolutions of \iris \citep{Pontieu2014_iris}.
We followed a procedure similar to that described in \cite{Pereira2013_iris_diagnostics_iii}; a Gaussian kernel of $0".4$ full width at half maximum (FWHM) is used for spatial degradation and a Gaussian profile with a FWHM of 6 pm for the spectral convolution. In addition, the spectra have been rebinned to a pixel size of $0".16 \times 0".33$ which translates to $199 \times 100$ pixels.

As will be described in Sect. \ref{sec:time-series} we synthesized a $10 \min$ time series at a cadence of $\approx 1.2 \,\mathrm{s}$ of the spectra emerging along artificial slits placed in the simulation domain. For each time step, the spectra are spatially degraded along the slits, but are not degraded across the slit as they are only one pixel wide. The spatial degradation is the same as described above: convolution by a Gaussian kernel of $0".4$ FWHM and binning to a pixel size along the slit of $0".16$.

%-------------------------------------------------------------------------------------------------------
\subsection{Identifying spectral features}                       \label{sec:methods_peak_identification}
%-------------------------------------------------------------------------------------------------------

\label{sec:peakfinder}
The typical shape of an observed \mgk line consists of two peaks that are commonly named \ktwov and \ktwor ("r":red, "v": violet/blue), a central minimum labeled as \kthree, and two minima in the "inner wings" of the spectral line labeled as \koner/\konev. The nomenclature for \mgh is similar. To classify the spectra, we need to identify characteristic spectral features such as the emission peaks and the \kthree feature. For this purpose we use a peak-finding algorithm that is part of the \texttt{ssw modules}\footnote{For an overview of IRIS related analysis tools see: \url{https://iris.lmsal.com/analysis.html}} \citep{ssw_1998SoPh..182..497F} and described in \cite{Pereira2013_iris_diagnostics_iii}. The synthetic spectra, however,  show the "classical" two-peaked profile shapes only in $\approx 20 \%$ of the cases (for \snapfourNineNine). The remaining $\approx 80 \%$ of spectra show three or more peaks. 
The identification of spectral features therefore has to be done carefully and comes with caveats.

After applying the peak-finding procedure we noticed the following hurdles.
First, the positions of the identified peak features appear to be correct in most of the cases. The corresponding intensities, however, are sometimes overestimated. This might be due to the peak-finding algorithm which interpolates the spectrum with a spline function to a regular grid, which sometimes may lead to an over- or under-shoot of the intensities. To overcome this issue we set the intensity of the identified features to the intensity without interpolation that is closest to the wavelength position of the feature.

Second, $\approx 3 \%$ of the identified peak features are found to be either both on the red, or both on the blue side of the spectrum, with respect of the central minimum.

Third, the dynamic nature of the atmosphere leads to spectra that have multiple peaks instead of only two. These can either be clearly distinguished, or they are close to each other in wavelength and thus appear as "wiggles" rather than as individual peaks. For those spectra even a visual characterization of the features is not straightforward and introduces some uncertainty.

In a typical snapshot we find that, before degrading the spectra, approximately $10$ \% of the peak features cannot be used as the peak-finding algorithm either failed (no results found) or delivered wrong results (\eg both \ktwo features on one side with respect to \kthree). Even after excluding these spectra from the statistics it turns out that profiles with multiple peaks may still be not correctly classified. This is because the wavelength position of a spectral feature in a profile with more than two peaks is not always uniquely defined from an observational point of view.

Degraded profiles are typically smoother, and the number of spectra with more than two peaks is significantly reduced. The peak features are therefore better defined and can be identified more clearly with the peak finder. A consequence of the smoother spectral profiles is also that some of the peaks become washed out, such that they appear as saddle points instead of minima/maxima. This occurs for 16 \% of the degraded spectra, for which the peak separation cannot be measured accurately.

%=======================================================================================================
\section{Results}                                                                    \label{sec:results}
%=======================================================================================================
In the following we discuss the properties of the synthetic spectra and compare them to the observation described in Sect. \ref{sec:methods_observations}. In Appendix \ref{sec:app-additional-observations} we compare the synthetic spectra with smaller selected FOVs and two additional observations to illustrate that the here presented average spectrum is representative for the purposes of this work. \mgh and k share similar formation properties and statistics \citep{Leenaarts_2013_mgii_hk_bifrost_diagnostics}. Therefore we focus here only on the \mgk line.
First, we show the spectra from snapshot \snapfourNineNine of the simulation and discuss it in detail. In Sect. \ref{sec:time-series} we compare this snapshot with three others. We then present in Sect. \ref{sec:discussion_muram_bifrost} a comparison with other numerical models.
The RT calculations in the selected snapshot converged in all but three of the columns, in which the code failed. In this analysis we focus on the average properties of the spectrum while in a future work we will study diagnostics that can be derived from spectral features of the intensity profiles, similarly to the works of \citet{Leenaarts_2013_mgii_hk_bifrost_diagnostics} and \citet{Pereira2013_iris_diagnostics_iii}.

% ++++++++++++++++++++++++++++++++++++++++++++++++++++++++++++++++++++++++++++++++++++++++++++++++++++
 \begin{figure*}
   \sidecaption
   \includegraphics[width=12cm]{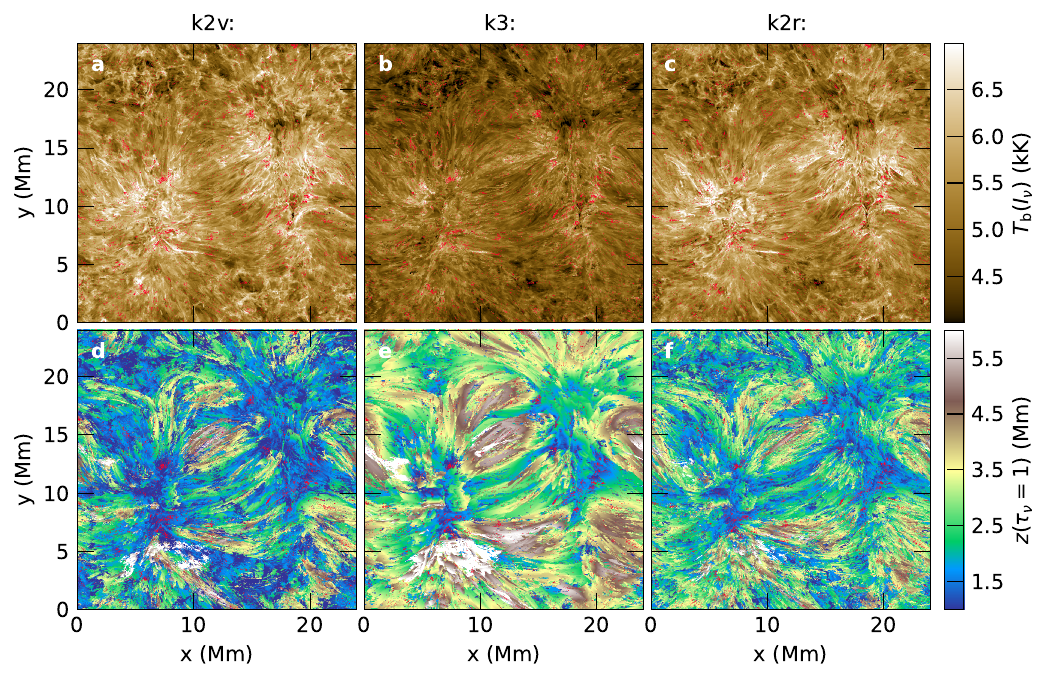}  
      \caption{Intensity and formation height at selected spectral features of the \mgk line. The top panels show the brightness temperature at \ktwov (panel a), \kthree (panel b), and \ktwor (panel c) as classified by the peak-finding algorithm. The bottom panels show the height at optical depth unity at the wavelength of \ktwov (panel d), \kthree (panel e), and \ktwor (panel f). Red colored pixels indicate where no feature could be detected.  The data corresponds to the non-degraded \snapfourNineNine snapshot.}        
      
         \label{fig:spectrograms_and_formation_heights}
   \end{figure*}
% ++++++++++++++++++++++++++++++++++++++++++++++++++++++++++++++++++++++++++++++++++++++++++++++++++++

%-------------------------------------------------------------------------------------------------------
\subsection{2D Images of \mgk}                                                 \label{sec:results_mgkLine}

In the following we present brightness temperature and formation heights of snapshot \snapfourNineNine measured at the \ktwo and \kthree features of the non-degraded synthetic spectra.

\subsubsection{Intensity maps}                 \label{sec:results_intensityMaps}
%-------------------------------------------------------------------------------------------------------

Figure \ref{fig:spectrograms_and_formation_heights}a-c show intensity maps of the spectral features \ktwov, \kthree, and \ktwor. The intensity is plotted in units of brightness temperature $T_\mathrm{b} = B^{-1}(I_\mathrm{\nu})$, where $I_\mathrm{\nu}$ is the emergent intensity and $B^{-1}$ the inverse Planck function. We emphasize that the conversion from $I_{\nu}$ to $T_\mathrm{b}$ is not linear and changes the statistical distribution and, in particular, the averages.
The intensity maps show structures that are well correlated with the magnetic field in the upper chromosphere (\cf Fig. \ref{fig:bzphotosphere}). 

In panels (d--f) it can be seen that the corresponding formation heights of the brightest structures are in the lower chromosphere ($<2 \Mm$). At these heights, the source function is only partially decoupled from the local atmospheric temperature leading to the brighter observed structures in comparison to other regions in the intensity images where the formation heights are higher up in the atmosphere with fainter brightness temperatures.

The upper and lower edges of the \ktwo intensity images  (Fig. \ref{fig:spectrograms_and_formation_heights}a,c) and to a lesser degree of the \kthree intensity image (panel b) show shock patterns associated with the quiet sun internetwork regions. Such enhanced shock patterns arise when wavefronts from different directions hit each other and appear similar to inverse granulation. These structures are less visible in the region between the two polarities (\ie roughly along the horizontal diagonal from the lower left to the upper right of the 2D images). This could be because the intensity at these locations forms in the connecting loops above the "lower chromosphere" (\cf panels d--f) which overlay the shock patterns. Moreover, in the region between the bipolar field concentrations the magnetic field is highly inclined, i.e. horizontal, which might suppress upward propagating shocks \citep[see, \eg,][]{gcauzzi_2007ASPC..368..127C_shocks}.

The appearance of the \ktwo intensity as brightness temperature looks qualitatively different from RT computations performed with the public Bifrost snapshot \citep[see, eg.][Fig.8]{Leenaarts_2013_mgii_hk_bifrost_diagnostics, sukhorukov2017}. The public Bifrost snapshot shows a larger amount of "shock expansion patterns" whereas in the \muram model the structures look more "fibril"-like. This suggests a higher opacity in the chromosphere and thus higher density compared with the \bifrost snapshot.

%-------------------------------------------------------------------------------------------------------
\subsubsection{Formation heights}                                \label{sec:results_formationHeightMaps}
%------------------------------------------------------------------------------------------------------

Figure \ref{fig:spectrograms_and_formation_heights}d--f show the formation heights of the \ktwov, \kthree, and \ktwor features. These are the heights where the optical depth at the corresponding wavelength reaches one. According to the Eddington-Barbier relation, in the case of optically thick line formation, this is approximately the height where the intensity at this wavelength is forming in the atmosphere. Similar to the intensities, the formation heights are spatially correlated with the magnetic field. The \ktwo peaks (panels d,f) form mostly below 3-3.5 Mm with exceptions where the plasma appears to form loop-like structures that extend higher up in the atmosphere. There, the formation heights are in the range $3.5$--$5 \Mm$. There are also regions where the line features form at even larger heights in the atmosphere such as visible, for example, in the lower left corner of panels (d)--(f) at roughly $(x,y)=(7 \Mm,5 \Mm) $.
It appears that, except in the magnetic loops, \ktwov forms a bit lower in the atmosphere than \ktwor.
The central minimum \kthree (panel e) forms in the higher part of the chromosphere just below the transition region. Overall, the variations in the \kthree image appear to be smoother than for the \ktwo peaks. This might partly be connected to misidentified peak features and therefore wrong formation heights. In the connecting loops visible in the center of the intensity images the \ktwo\ peaks show too high formation heights (higher than \kthree). This is because the profiles associated with these rays have multiple peaks and the classical features of the \mgk line are therefore not well defined. It might therefore be that the selected \kthree feature forms at lower heights than the selected \ktwo features.

%-------------------------------------------------------------------------------------------------------
\subsection{Comparison with IRIS observations}                                    \label{sec:results_averageSpectra}
%-------------------------------------------------------------------------------------------------------

We now discuss the spatially-averaged spectrum that we compute over the full simulation domain. We compare this to a spatially-averaged \iris raster scan which is chosen to include quiet Sun network and internetwork regions, similar to those seen in our simulations. However, the field topology in the observations is different, and the FOV is larger. The mean unsigned LOS magnetic field strength averaged over the shown FOV in the HMI magnetogram is $\langle|B_{\mathrm{LOS}}/\mu| \rangle=12.12 \G$. This is lower than the vertical component of the magnetic field estimated from the simulation at $\tau_{500}=0.1$, which was $\approx 20 \G$ (after degrading to the same resolution, see Sect. \ref{sec:methods_enhanced_network_model}).

\begin{figure*}
   \centering
  
   \includegraphics[width=16.4cm,clip]{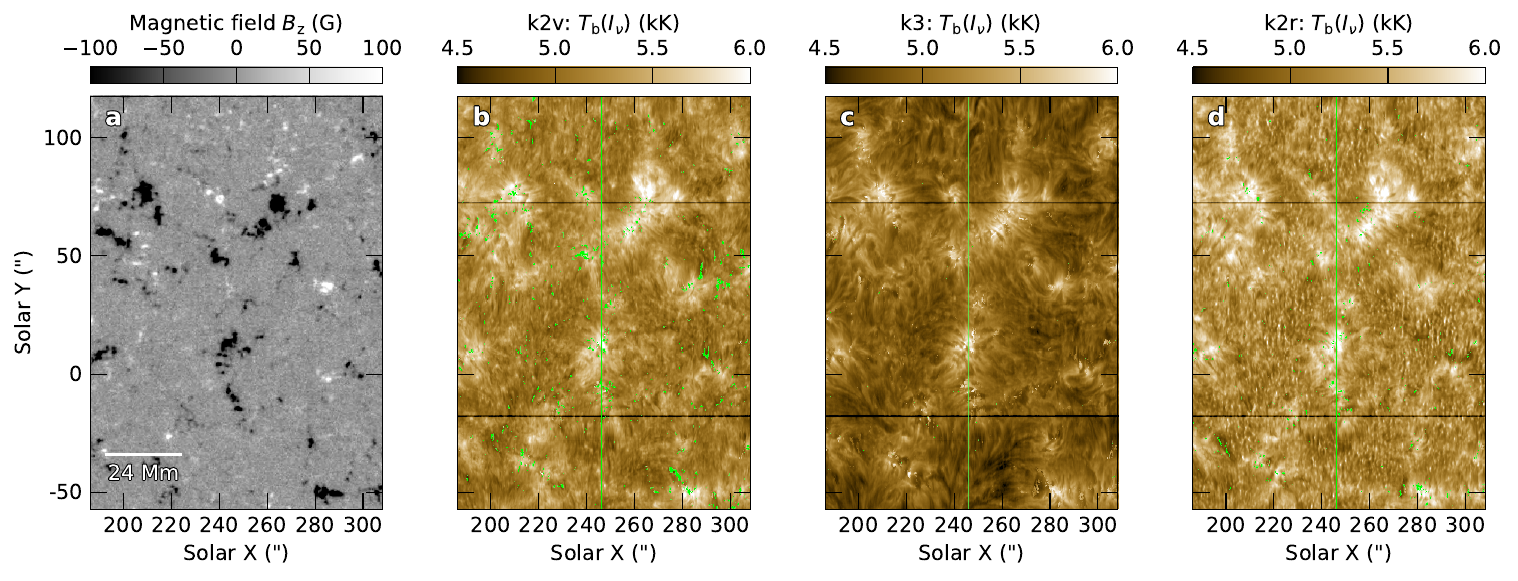}
    
      \caption{Observations from a quiet sun region including network fields. The left panel shows the HMI magnetogram that was aligned with the IRIS observation. The gray scale in the \hmi magnetogram is clipped to $\pm 100 \mathrm{\, G}$ but values up to $\pm 1000 \mathrm{\, G}$ are reached. Panels (b)--(d) show the observed intensity as brightness temperature $T_{\mathrm{b}}$ at the \ktwo and \kthree features of the \mgk line. The limits of the colorscale are not the same as in Fig. \ref{fig:spectrograms_and_formation_heights} in order to increase the contrast. The size bar in panel (a) indicates the size of the simulation domain. Green colored pixels indicate where no features could be found or no measurements are available.}
         \label{fig:comparison_observation}
   \end{figure*}

 The full FOV of the observations chosen for the comparison is shown in Fig. \ref{fig:comparison_observation}. The LOS magnetic field map (panel a) shows features of both polarities that are distributed over the whole FOV. At $(x,y)\approx (270",45")$ a bipolar structure is visible that qualitatively compares to the setup of our simulation. As described in Appendix \ref{sec:app-additional-observations} the properties of a smaller ROI around the bipole and the full FOV are similar.
  
 The intensity images of the spectral features are extracted by applying the peak-finding algorithm described in Sect. \ref{sec:peakfinder}. We show the \ktwov, \kthree, and \ktwor features in brightness temperature units. A similar correlation is found between the strong network field regions and the intensities as in the model (\cf Figures \ref{fig:bzphotosphere},\ref{fig:spectrograms_and_formation_heights}). The intensity in the \ktwo peaks (panels b and c) and the central reversal is enhanced in the presence of magnetic features. The overall contrast of the observations appears to be smaller than in the simulation. This is expected from the lack of horizontal radiative transfer in our synthetic spectra \citep{sukhorukov2017}. We note that the data shown in Fig. \ref{fig:spectrograms_and_formation_heights} is not degraded to preserve the information about the formation height.

% ++++++++++++++++++++++++++++++++++++++++++++++++++++++++++++++++++++++++++++++++++++++++++++++++++++
% ++++++++++++++++++++++++++++++++++++++++++++++++++++++++++++++++++++++++++++++++++++++++++++++++++++

% ++++++++++++++++++++++++++++++++++++++++++++++++++++++++++++++++++++++++++++++++++++++++++++++++++++
  \begin{figure*}
   \centering
     \includegraphics[width=\hsize,clip]{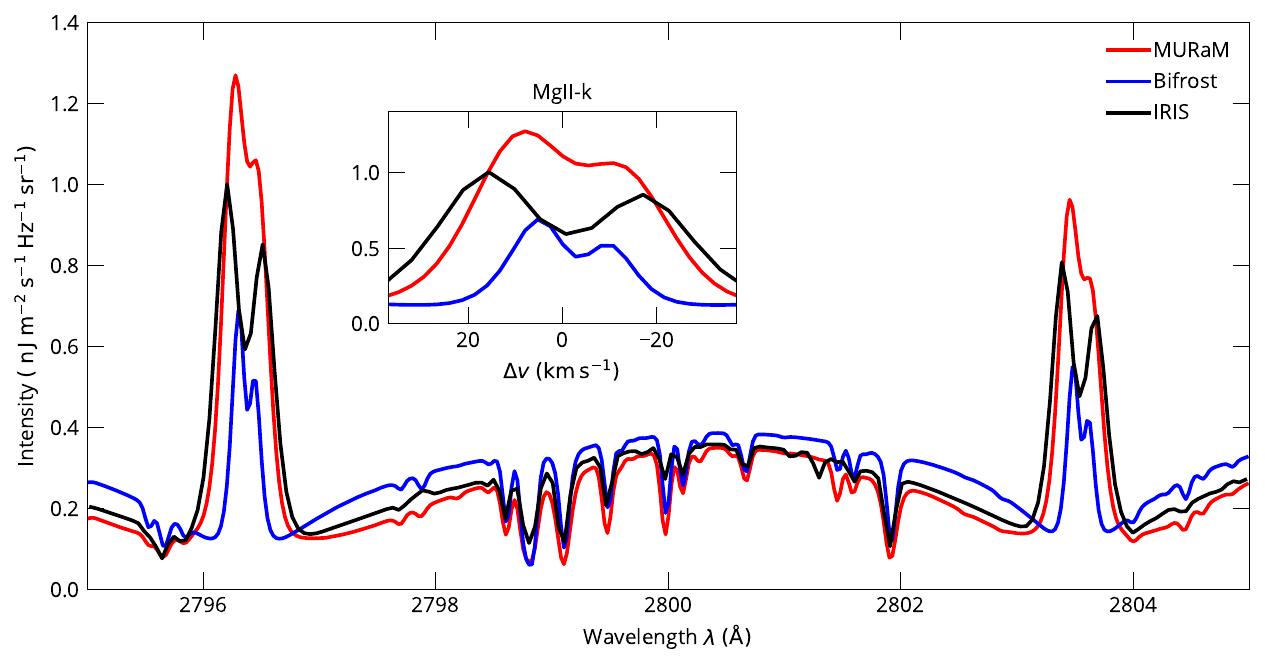}

      \caption{Spectrum of \mghk lines. Shown are spatial averages of the spectra observed by IRIS (black) and synthesized from snapshot \snapfourNineNine (red). For comparison, we added the publicly available spectrum from \bifrost (blue). A comparison between the two synthetic spectra is presented in Sect. \ref{sec:discussion_muram_bifrost}. The inset panel shows a zoom to the \mgk line as a function of Doppler velocity centered on the rest wavelength.}
         \label{fig:av_spectra}
   \end{figure*}
% ++++++++++++++++++++++++++++++++++++++++++++++++++++++++++++++++++++++++++++++++++++++++++++++++++++

%-------------------------------------------------------------------------------------------------------
\subsubsection{Average spectra}              \label{sec:results_averageSpectra_comparisonToObservationsAverageSpectrum}
%-------------------------------------------------------------------------------------------------------
In Fig. \ref{fig:av_spectra} we present average spectra from a representative \iris observation, the \muram  model, and the \bifrost publicly available snapshot. The synthetic spectra were degraded to the \iris spectral and spatial resolution using the procedure described in Section~\ref{sec:methods_degradation}. In the following we discuss the shape of the profiles in terms of intensities and peak separation. 

The intensities at \kone and in the wings of the observed and \muram spectra are similar. However, the observed intensity appears on average to be slightly stronger there. The intensities at the \ktwor and \ktwov peaks are higher in the \muram model compared to the observations. Some of the differences might be attributable to the 1.5D RT treatment and some to the discrepancies between the observation and the modeled region. A comparison between Fig. \ref{fig:bzphotosphere} and Fig. \ref{fig:spectrograms_and_formation_heights}a,c shows that the brightness temperature is clearly enhanced above the network magnetic field elements. The high brightness temperatures ($T_{\mathrm{b,k_2}}>5 \, \mathrm{kK}$) may be explained by high temperatures in the atmosphere at the corresponding formation heights \citep{Leenaarts_2013_mgii_hk_bifrost_diagnostics}. \citet{sukhorukov2017} found, by studying the public \bifrost snapshot in 3D RT, that brightness temperatures in the peaks with $T_{\mathrm{b,k_2}} > 5 \, \mathrm{kK}$ are overestimated in the 1.5D RT computation. We therefore expect a reduction of the \ktwo peaks in the intensity images as well as in the average spectrum in a 3D RT computation.

Both the observed and the simulated spectrum have, on average, stronger intensities in the \ktwov peaks. We computed the intensity ratio $R_\mathrm{k} = (I_{\mathrm{k2v}} - I_{\mathrm{k2r}}) / (I_{\mathrm{k2v}} + I_{\mathrm{k2r}})$ as in \citet{Leenaarts_2013_mgii_hk_bifrost_diagnostics}, where the intensities $I$ are measured at the blue and red peaks of the spatially averaged spectrum. The obtained values of $0.09$ (\muram) and $0.08$ (\iris) are both positive and similar. According to \cite{Leenaarts_2013_mgii_hk_bifrost_diagnostics} there is a correlation between $R_\mathrm{k}$ and the sign of the average velocity in the atmosphere between the formation height of the \ktwo and \kthree features. In the analyzed snapshot we find $56\%$ of the columns show on average a downflow and $31\%$ an upflow. For the remaining columns, no such correlation with the average velocity could be determined  because of erroneous information about the formation height, as a consequence of complex profile shapes.

The intensity at the central reversal \kthree is larger in the averaged \muram spectrum. It turns out that this is an effect of superposition of single spectra when the average is computed. 
The intensity at \kthree in the individual profiles identified by the peak finder is, on average, about $20 \%$ smaller than \kthree intensity in the spatially averaged profile.
As shown in \cite{Leenaarts_2013_mgii_hk_bifrost_diagnostics} the Doppler shift of the central reversal correlates with the velocity at the corresponding formation height. Therefore, spectra from rays with large velocities in the upper chromosphere will influence the shape and the intensity of the central reversal. Whether this effect is similarly dominant in 3D RT spectra will be investigated in a future work.

%%%%%%%%%%%%%%%%%%%%%%%%%%%%%%%%%%%%%%%%%%%%%%%%%%%%%%%%%%%%%%%%%%%%%%%%%%%%%%%%%%%%%%%%%%%%%%%%%%%
\subsubsection{Distribution of peak intensities}
%%%%%%%%%%%%%%%%%%%%%%%%%%%%%%%%%%%%%%%%%%%%%%%%%%%%%%%%%%%%%%%%%%%%%%%%%%%%%%%%%%%%%%%%%%%%%%%%%%%

\begin{figure}
   \centering
   \includegraphics[width=\hsize,clip]{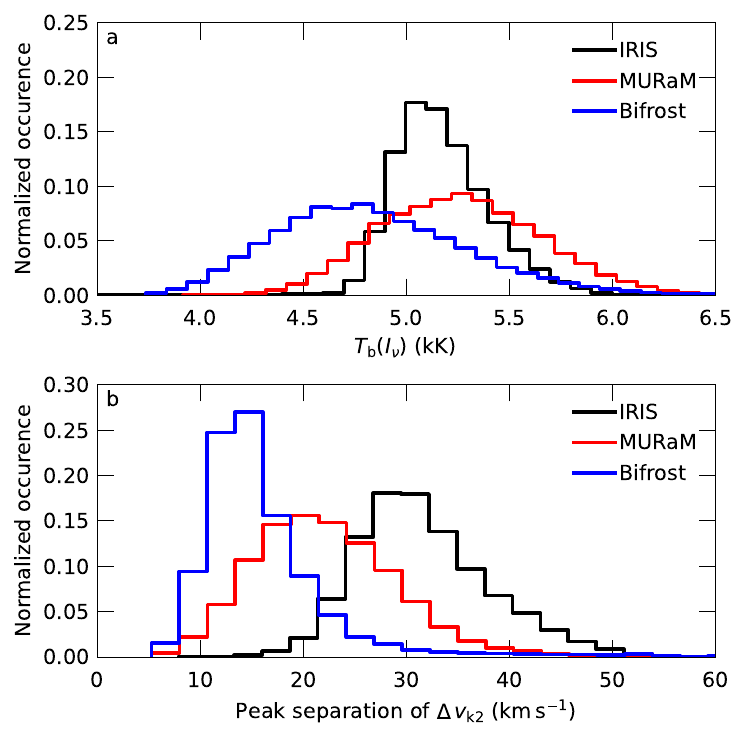}
      \caption{Distribution of \ktwo peak intensities in units of brightness temperature (panel a) and peak separations (panel b) of the \mgk line for observations from \iris (black) and the enhanced network model from \muram (red). For comparison, histograms of the same line parameters obtained from the \bifrost public snapshot have been over-plotted in blue (see Sect. \ref{sec:discussion_muram_bifrost} for a discussion). For these statistics we used the modeled spectra that are degraded to the observed spatial and spectral resolution for a meaningful comparison.}
         \label{fig:peak_separation}
   \end{figure}

The spatially averaged spectrum is smooth. Spectra from individual rays, however, show large variations in their shape. Many profiles show more than two emission peaks with a range of intensities. We therefore investigate how peak intensities and separations are distributed following the example given in Fig. 7 of \cite{carlsson_chromosphere_review2019}. For this purpose we use only spectra that were degraded to reflect \iris instrument conditions (see Sect. \ref{sec:methods_degradation}). In Fig. \ref{fig:peak_separation}a it can be seen that the computed peak brightness temperature distribution covers all observed brightness temperatures and also extends to temperatures that are about 500 K lower and 500 K higher than observed. Overall the width of the observed distribution is smaller than that of the computed distribution. The shown distribution from the observation contains the full FOV. In Appendix \ref{sec:app-additional-observations} we show that similar distributions can be found in smaller selections of the FOV that are comparable in size to our simulation, given the network fields inside the region of interest are not too strong. We find that a distribution similar to the one in \citet{carlsson_chromosphere_review2019} can be found by choosing a quiet sub-region of the observation (Appendix \ref{sec:app-additional-observations} 'Quiet sun'), which is basically devoid of all network features.  The distribution presented in the paper (Fig. \ref{fig:peak_separation}) is more representative of observations which include regions of stronger magnetic field.

% ++++++++++++++++++++++++++++++++++++++++++++++++++++++++++++++++++++++++++++++++++++++++++++++++++++
     
% ++++++++++++++++++++++++++++++++++++++++++++++++++++++++++++++++++++++++++++++++++++++++++++++++++++

\subsubsection{Peak separation}

\label{sec:resultsPeakSeparations}
% ++++++++++++++++++++++++++++++++++++++++++++++++++++++++++++++++++++++++++++++++++++++++++++++++++++
  \begin{figure}
   \centering
   \includegraphics[width=\hsize,clip]{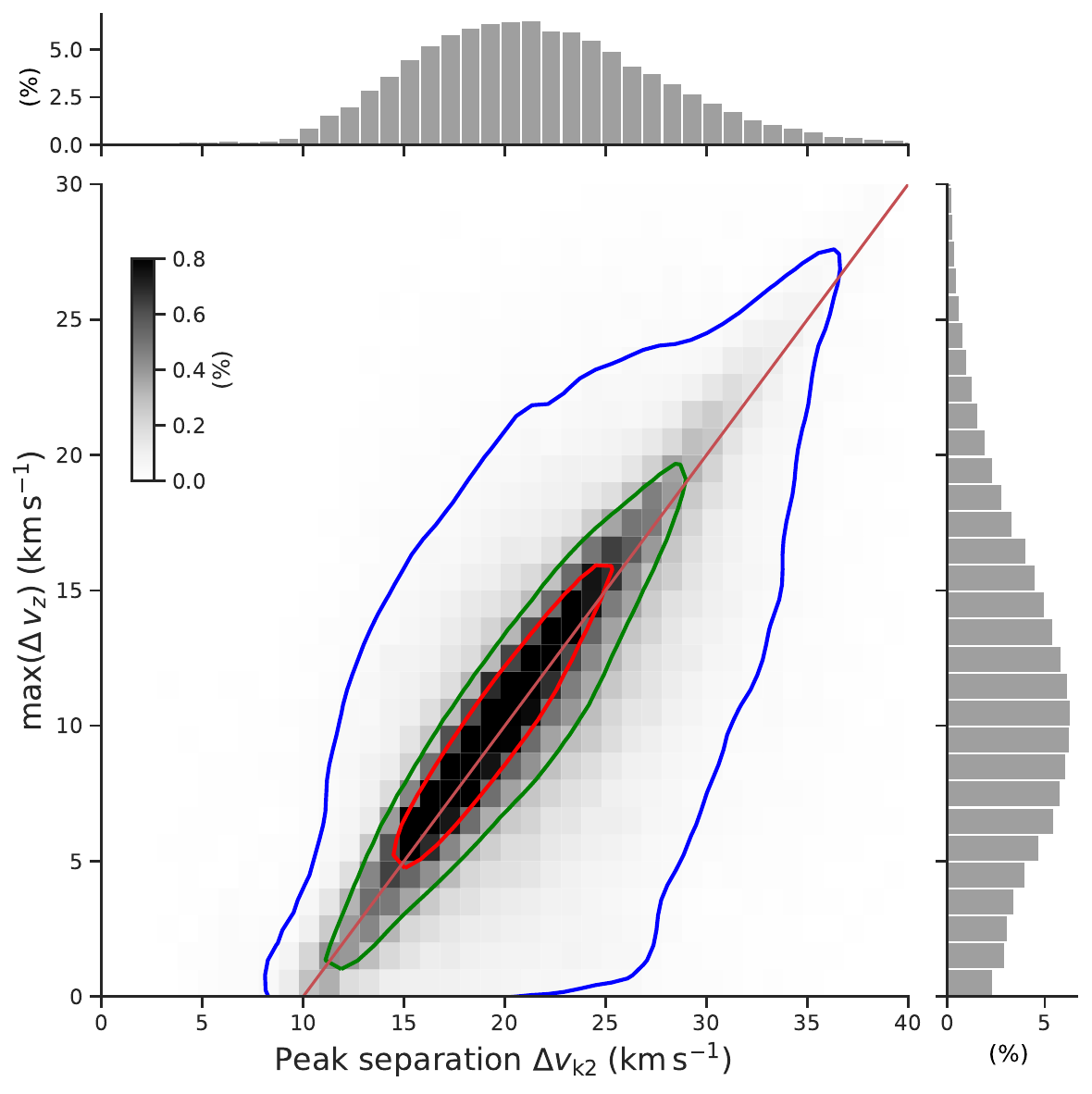}

      \caption{Correlation between peak separation and maximum velocity difference measured from double peaked spectra. The peak separation is defined as the difference between the wavelengths of \ktwov and \ktwor in units of the Doppler velocity. The maximum velocity difference, $\mathrm{max(\Delta v_{z })}$, is defined as the difference between the maximum and minimum velocity in the atmosphere between the formation height of the central minimum \kthree and the minimum of the of the \ktwor and \ktwov formation heights. The blue, green, and red contours enclose $90 \%,50\%,$ and $25\%$ of the data set, respectively. Both quantities are measured from the not degraded spectra. The gray scale of the pixels in the 2D histogram is clipped to $0.8\%$ to increase the readability. The top and right panels show the marginal distributions in percentage. }
         \label{fig:peak-separation-vgradient}
   \end{figure}
% ++++++++++++++++++++++++++++++++++++++++++++++++++++++++++++++++++++++++++++++++++++++++++++++++++++

In comparison to the line width (see Sect. \ref{sec:lineWidth}), which is not uniquely defined, the peak separation is straightforward to measure once the spectral features are identified. We measured the peak separation of the degraded synthetic spectra and the observations. The distributions of the peak separation is shown in Fig. \ref{fig:peak_separation}b. It can be seen that the observed spectra peak around $30 \kms$ with an average value of $\langle\Delta v_{\mathrm{k2}}\rangle = 33.02  \kms $, whereas the distribution of the peak separation from \muram profiles peaks slightly above $20 \kms$ and are on average $\langle\Delta v_{\mathrm{k2}}\rangle = 23.6 \kms $. The most likely values for the peaks separation are by $8$--$10 \kms$ smaller than in the observed spectra.

The distribution of the peak separation is sensitive to the degradation procedure. This occurs because degrading the profiles to \iris resolution reduces the number of small peaks. The decreased complexity of the profiles makes it easier to define and identify the \ktwo features. After degrading the synthetic profiles, the peak of the distribution changes from $\approx 16 \kms$ to $\approx 20 \kms$.

\cite{Leenaarts_2013_mgii_hk_bifrost_diagnostics} showed that the peak separation strongly correlates with velocity differences in the mid to upper chromosphere. In order to understand the peak separation in the \muram model, we tested whether a similar correlation can be found. We therefore measured in a similar manner to \citet{Leenaarts_2013_mgii_hk_bifrost_diagnostics} for each atmospheric column the maximum velocity difference 
\begin{align}
    \mathrm{max}(\Delta v_{z}) &= \max_{z_1\leq z \leq z_2}(v_z) - \min_{z_1\leq z \leq z_2}(v_z),
    \label{eq:maxdv}
\end{align}
with
\begin{align}
        z_1 &=\min (  z(\tau_{\mathrm{k2r}}=1) , z(\tau_{\mathrm{k2v}} = 1) ), \\
        z_2 &= z(\tau_{\mathrm{k3}}=1),
\end{align}
in the atmosphere between the minimum formation height of the (non-degraded) \ktwo peaks and the \kthree feature, respectively. We note that our definition of $z_1$ is different. We chose this definition because in the \muram atmosphere the formation height of \ktwor can be different by up to $1$--$3 \Mm$ from that of \ktwov. In addition, we restricted the measurement of $\mathrm{max}(\Delta v_{z})$ only to locations in the atmosphere where the plasma temperature is $T_{\mathrm{gas}} \leq 10 \, \mathrm{kK}$ \citep{carlsson_2012_approximations}. The latter condition was necessary as there are columns in the atmosphere where \kthree forms much higher than the \ktwo features with hot plasma in between.
A qualitative comparison between a vertical slice in temperature of our model (Sect. \ref{sec:lineWidth}) and the Bifrost snapshot \citep[\eg][Fig. 4]{carlsson_chromosphere_review2019} suggests that the public \bifrost model has a less corrugated transition region in the upper chromosphere. It might therefore be that these scenarios are much less present in the public \bifrost model explaining the above-mentioned differences. The study of \citet{trujilo_bueno_2018ApJ...866L..15T}, based on comparisons to the linear polarization signal measured by the Chromospheric Lyman-Alpha SpectroPolarimeter (CLASP) rocket experiment \citep{kano_clasp_2017ApJ...839L..10K}, also concluded that the Bifrost model required additional corrugation at the transition region to match the observations.
For $15.2 \%$ of the columns either no peak separation or no $\mathrm{max}(\Delta v_{z})$ could be determined because of missing or wrong information about the formation height which is due to miss identification of the peak finder. The data used for the correlation therefore corresponds to $84.8 \%$ of the columns in the analyzed snapshot.  We find a positive correlation between the peak separation and $\mathrm{max}(\Delta v_{z})$ with a correlation coefficient of $0.48$. As mentioned above, the \ktwo separation is difficult to measure for \mgk profiles with multiple peaks. We therefore measured the correlation again, but preselected spectra showing only two peaks. We then found a larger correlation coefficient of $0.59$, which is comparable to that reported by \citet{Leenaarts_2013_mgii_hk_bifrost_diagnostics}. The correlation for spectra with two peaks is shown in Fig. \ref{fig:peak-separation-vgradient}. It can clearly be seen that for the majority of the spectra with an average peak separation of $\approx 20 \kms$, $\mathrm{max}(\Delta v_{z})$ values of $10$--$15 \kms$ in the corresponding columns of the atmosphere are needed.
For spectra that do not clearly follow the correlation, the peak separation is larger than $\mathrm{max}(\Delta v_{z })$. This might be a consequence of a local maximum in the temperature in the lower atmosphere, possibly as a result of a propagating shock. This leads to an increased peak separation unrelated to the velocity profile \citep[see also Fig. 8a,b in ][and the discussion therein]{Leenaarts_2013_mgii_hk_bifrost_diagnostics}. We discuss the role of the velocity field in the broadening of the \mgk line in Sect. \ref{sec:lineWidth}.

%----------------------------------------------------------------------------------
\subsubsection{Line width}                             \label{sec:lineWidth}
%----------------------------------------------------------------------------------

\begin{figure*}
\sidecaption
\includegraphics[width=12cm]{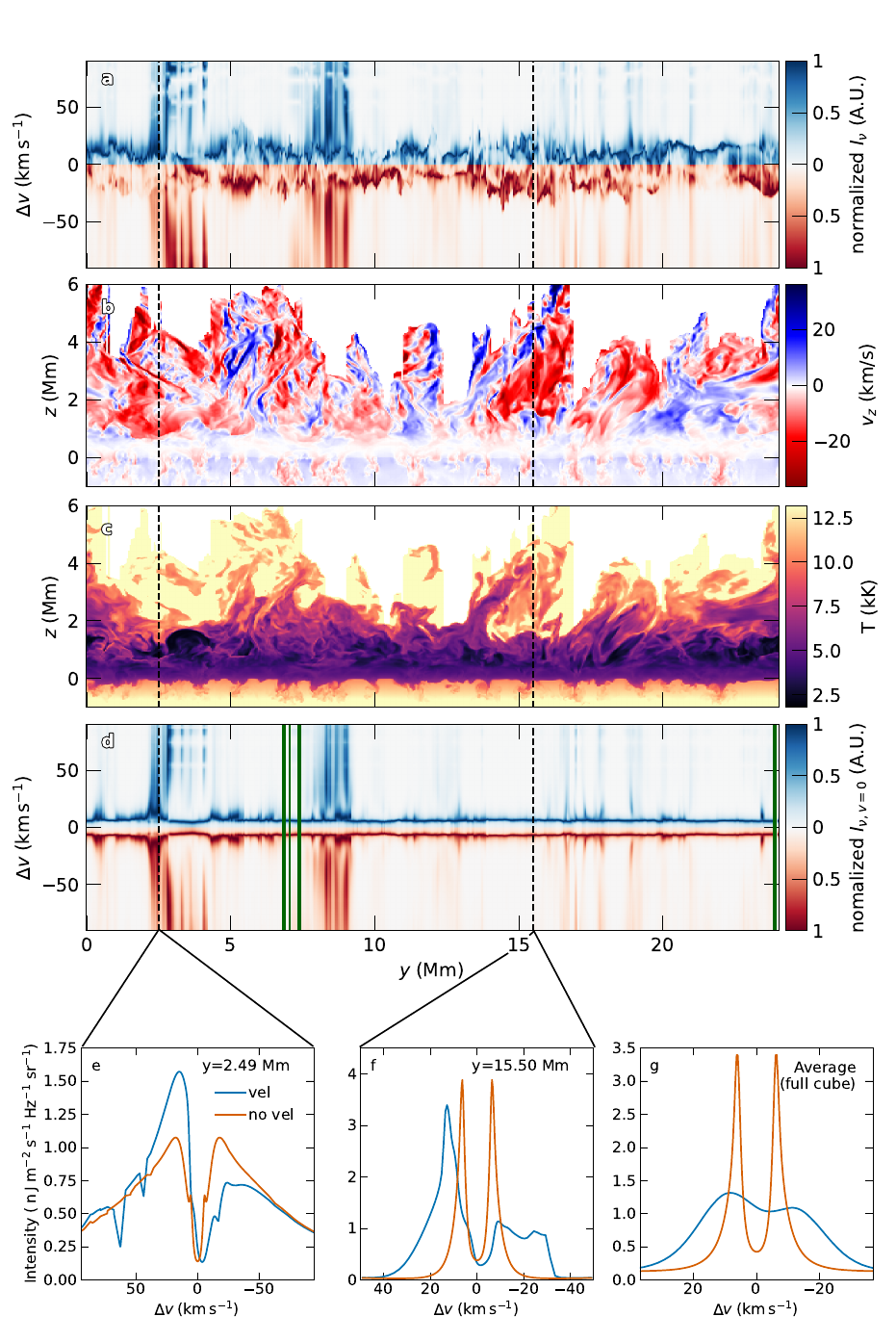}
  \caption{Contribution from velocities to emergent spectra along a cut at $x=13.95 \Mm$ through the \muram snapshot \snapfourNineNine. Panel (a) shows the spectrum of the \mgk line with color coded intensity. Red colors refer to the red part of the spectrum and blue colors to the blue/violet part of the spectrum relative to the rest wavelength of \mgk. The intensities are normalized to the maximum value in the presented range of $-90 \leq \Delta \,v / [\kms] \leq 90$. Panel (b) shows the vertical velocity component and panel (c) the temperature in the atmosphere. Panel (d) is similar to panel (a) but for a RT computation of the identical snapshot but neglecting velocities in the atmosphere. The green colored regions in panel (d) indicate columns where the RT computation failed. 
Panels (b,c) show only the part of the atmosphere that was used for the RT computation (see Sect. \ref{sec:methods_synthesis}). The bottom panels (e) and (f) show comparisons of single \mgk spectra at $y=2.49 \Mm$ and $y=15.5 \Mm$ between the computation with and without velocities. In panel (g) we show a comparison of the globally averaged spectrum between the two computations. We note that neither the y-axis nor the x-axis are the same for panels (e)--(g).
  }
     \label{fig:temp-spec-source-function}
\end{figure*}

In this Sect. we demonstrate that the width of the emergent spectra in the \muram atmosphere is mainly determined by the velocities.

We study the impact of velocities on the emergent spectra by solving the RT problem in snapshot \snapfourNineNine but neglecting the velocity field. We emphasize that in this comparison only the "direct" effects of the velocities in the RT computation are neglected. The structure of the temperature profile which is used in both computations as well as all other physical quantities are a result of the MHD computation from which the velocities cannot be separated. In particular, local maxima in temperature in the lower atmosphere are often a result of compression heating by propagating plasma waves. In the RT computation without velocities we found that the RT computation failed for $0.6\%$ of the rays. The missing spectra, however, have no influence on the conclusions drawn in this section.

In Fig. \ref{fig:temp-spec-source-function} we show a comparison of the results along one cut in the $y$-axis at $x=13.95 \Mm$ of the atmosphere. Panels (a) and (d) show the intensity of the \mgk line along the spatial axis with and without velocities. For a better understanding, the intensities on the blue side of the line's rest wavelength are shown in blue and the corresponding red part of the spectrum in red. Both color scales are normalized to the maximum intensity in the shown wavelength range. 
Panels (b) and (c) show the vertical velocity and the temperature to show the context of the underlying atmosphere. We define upflows as positive velocities in the atmosphere.

As expected, the spectra computed with velocity (panel a) display more variability in space (and time -- see Sect. \ref{sec:time-series}) and are significantly broader than those computed without velocity (panel d). The latter are symmetric around the rest wavelength of \mgk and the variation of the line width in space seems much smaller. Qualitatively, it can be seen that the broadest spectral lines appear at locations where the temperature has peaks in the lower atmosphere (e.g., panel (c) at $\approx 7-9 \Mm$). At around $3-5 \Mm$ and $7-9 \Mm$ the line widths are larger than the average width in both cases. As an example we show the spectrum for the same ray from both computations in panel (e). It can be seen that the \ktwo intensity and the peak asymmetry is different in the "velocity" case. The wavelength position and the overall line width is similar in both cases. 
In the above-mentioned regions which have broad line profiles in both cases, a local maximum in temperature in the lower atmosphere leads to the larger line width.

In panel (f) we compare the spectra from the same ray for the two calculations at the location where the line width is different. As in panel (e) the spectrum computed without the velocity field is symmetric. However, the line width and the peak separation in the "velocity" case is much larger. The larger line width is here a result of a broader extinction profile as a consequence of large velocity variation along the line of sight. To demonstrate that the line width in the spatially averaged spectrum is dominated by the influence of velocities, we show in panel (g) the average spectrum over the whole simulation domain from \snapfourNineNine with and without velocities. We computed the averages only over rays that converged in computations both with and without the velocity field. There is, however, no visible difference between the blue line in panel (g) and the red line in Fig. \ref{fig:av_spectra}, because the missing $0.6\%$ of the rays have no impact on the overall average.

We conclude from this study that, in the case of the \muram model, while local temperature maxima in the lower atmosphere can have a significant impact on the width of resulting spectral line profiles, in the majority of the computed spectra the line width is dominated by the velocities in the upper chromosphere.

% ---------------------------------------------------------------------------------------------------
\subsubsection{Time variation}                                             \label{sec:time-series}
% ---------------------------------------------------------------------------------------------------

  \begin{figure}
   \centering

   \includegraphics[width=\hsize,clip]{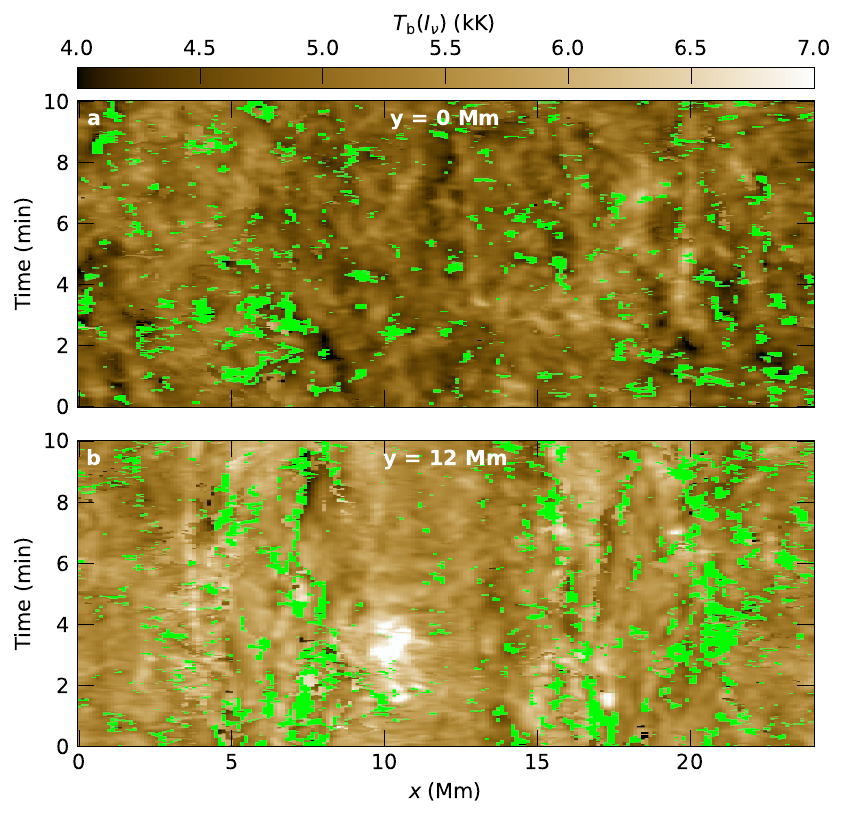}

      \caption{Space-time intensity maps. The panels show the time variation (vertical axis) of brightness temperature measured at \ktwor along two spatial slits (horizontal axis) located at the positions $y=0 \Mm$ (panel a) and 12~Mm (panel b). The spectra were degraded before the peak-finding algorithm was applied. Panel (a) represents the quieter regions of the simulation domain. Panel (b) represents a cut through the network dominated regions of the simulation. The color-scale is set to the same limits as Fig. \ref{fig:spectrograms_and_formation_heights}. Green pixels indicate spectra where no \ktwor feature could be detected.}
         \label{fig:time-intensity-slits-overview}
   \end{figure}

The results discussed so far are from a single snapshot in the simulation which corresponds to one time step in the computation. The conditions in the solar atmosphere are constantly changing with time and it is not immediately clear whether a spatial average can account for temporal variation of the line.

In order to get at least a very rough idea of how temporal effects in our simulation affect the \mghk profiles, we synthesized three more snapshots that are approximately separated by 2 minutes each. For these snapshots we synthesized every second column in $x$ and $y$ direction (see Sect. \ref{sec:methods_synthesis}). Considering a larger number of "full" snapshots was not feasible because of computational costs. To get better insight into how the spectral lines evolve with time, we  additionally chose 8 "synthetic slits" in the simulation located at $x,y \in [0,6,12,18] \Mm$, oriented in the $x$ and $y$ directions. The time resolution of the profiles computed along the synthetic slits is approximately $1.2 \sec$. Figure~\ref{fig:time-intensity-slits-overview} shows as an example the time variation for two slits that are located at $y=[0,12] \Mm$. In the quieter part of the simulation box (represented by panel (a) of Fig.~\ref{fig:time-intensity-slits-overview}; \cf Fig. \ref{fig:bzphotosphere} at $y=0 \Mm$) the brightness temperature measured at the \ktwo feature shows similar shock patterns as visible in Fig. \ref{fig:spectrograms_and_formation_heights}a. Panel (b) represents the network-dominated part of the simulation domain. Here, the brightness temperature is on average $\approx 630 \, \mathrm{K}$ higher than in panel (a). While panel (a) shows repeating shock-like patterns, panel (b) shows more structures with a larger extent along the spatial axis, which come from the connecting loops between the two polarities. Both slits contain also brighter features that live for more than $4 \min$ as, for example, panel (a) at $x\approx 20 \Mm$ and panel (b) at $x\approx 4 \Mm$.

     \begin{figure}
   \centering
   \includegraphics[width=\hsize,clip]{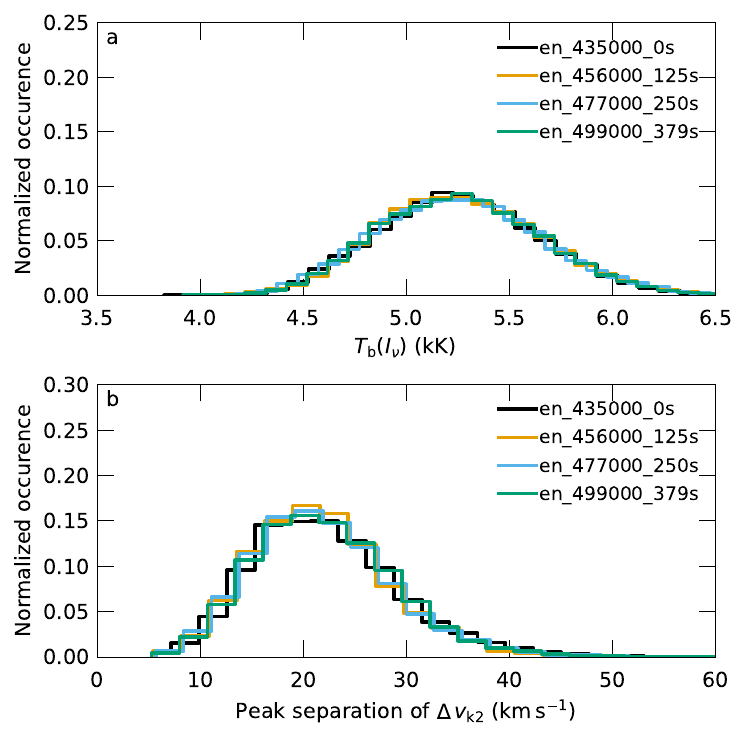}
      \caption{Distribution of \ktwo peak intensities in units of brightness temperature (panel a) and peak separation (panel b) for four snapshots in the \muram simulation which are separated by $\approx 2 \min$ of simulation time. 
      The presented data corresponds to every second column in $x$ and $y$ direction in the snapshots. The spectra were degraded before the peak-finding algorithm was applied. For comparison with Fig. \ref{fig:peak_separation} we chose the same limits for the axes.
      }
         \label{fig:peak_separation_all_whole_snaps}
   \end{figure}

  \begin{figure}
   \centering

   \includegraphics[width=\hsize,clip]{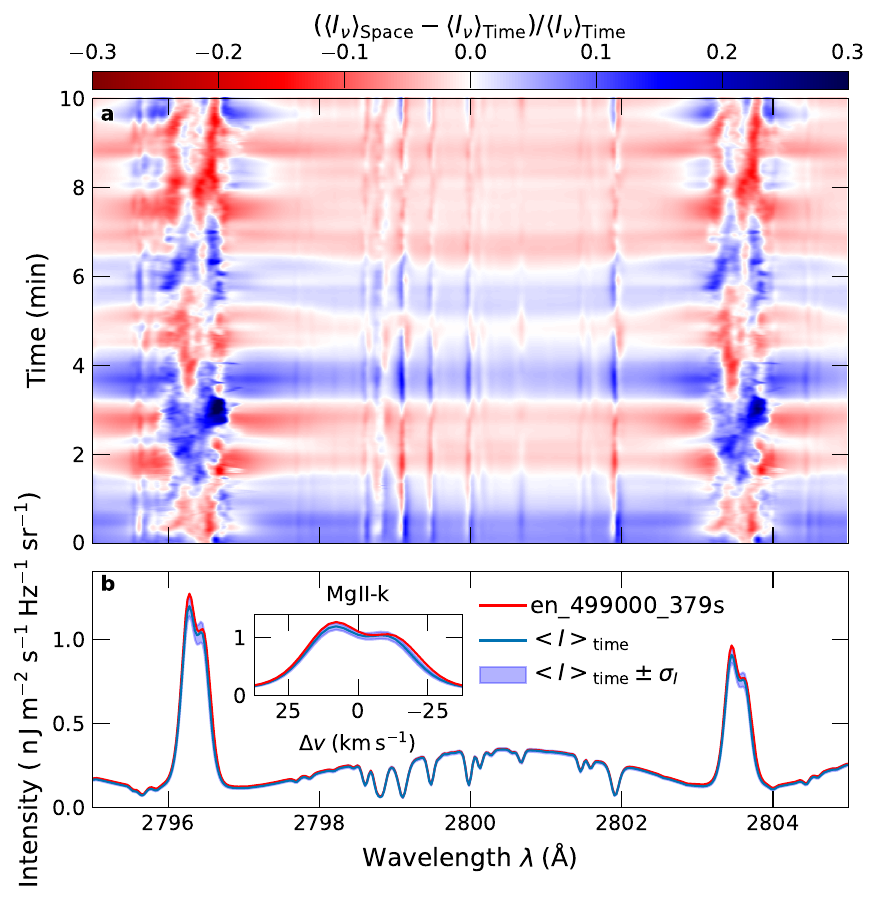}

      \caption{Time variation of the spatially averaged \mghk spectrum. Panel (a) shows a time series of $10 \min$ length at a cadence of $\approx 1.2 \sec$. We computed the spatial average from 8 slits that are evenly spaced in the simulation at positions $x,y \in [0,6,12,18] \Mm$. We then subtracted the average spectrum over the whole $10 \min$ and show the relative difference. In Panel (b) we show the time averaged spectrum together with the standard deviation. The red curve shows for comparison the spatially averaged spectrum from snapshot \snapfourNineNine that was discussed in detail in Sect. \ref{sec:results_averageSpectra}. The spectra have been degraded to the IRIS spatial and spectral resolution.}

            \label{fig:av-spectrum-ts-time-variation}
   \end{figure}

While the spectra can change significantly over time on small spatial scales, the average and statistical properties are roughly constant. To demonstrate this we show in Fig. \ref{fig:peak_separation_all_whole_snaps} the distribution of peak intensities and peak separation for the 4 snapshots at different times. There are no significant changes over time in a statistical sense. The average values of the peak separations range from $\langle\Delta v_{\mathrm{k2}}\rangle = 23.1\kms $ to $\langle\Delta v_{\mathrm{k2}}\rangle = 23.6 \kms$. 

We use the spectra from the synthetic slits to estimate the spatially averaged spectrum. This is demonstrated in Fig. \ref{fig:av-spectrum-ts-time-variation}. We calculated the spatial averages over all slits at each time step, which we call $\langle I_{\nu} \rangle|_{\mathrm{Space}}$. We then computed the temporal average over all time steps (\ie all  $\langle I_{\nu} \rangle|_{\mathrm{Space}}$) which we call  $\langle I_{\nu} \rangle|_{\mathrm{Time}}$ and subtracted this from each single spatially averaged spectrum.
In panel (a), it can be seen that the estimated average spectrum shows variations with time that are close to the three-minute period usually prominent in the chromosphere. Whether it is really periodic and what its periodicity is cannot be inferred from this short time series. 
In panel (b) the time-averaged spectrum is shown together with the standard deviation of the different time samples. It can be seen that at the wavelength position of the \ktwo (and \htwo) features the standard deviation is largest. This is expected because these features of the spectral lines form in the mid to upper chromosphere where the atmosphere is more turbulent compared to lower layers. Nonetheless, the deviation from the time average  is relatively small, as is the difference 
to the snapshot \snapfourNineNine discussed above.

\subsection{Comparison to other numerical models}
\label{sec:discussion_muram_bifrost}

The public \bifrost snapshot \citep{Carlsson2016a_public_bifrost_snapshot} has become the standard for analyzing spectral line formation in the solar chromosphere over the last years. In this Sect. we briefly compare the  synthetic spectra resulting from our \muram simulation run and the \bifrost public snapshot as well as the average physical properties of the two modeled atmospheres. 

% ++++++++++++++++++++++++++++++++++++++++++++++++++++++++++++++++++++++++++++++++++++++++++++++++++++
\begin{figure}
   \includegraphics[width=\hsize,clip]{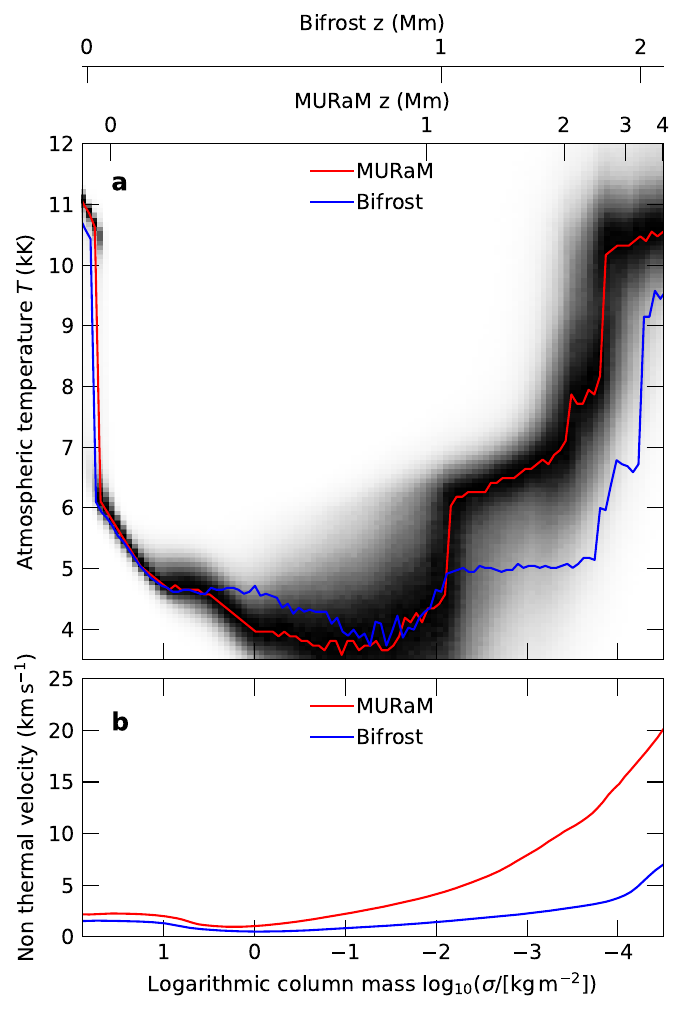}
      \caption{Comparison between atmospheric properties between the \muram and the \bifrost public snapshot. Panel (a): Atmospheric temperature as a function of column mass $\sigma$. The two curves show the peak of the temperature distribution of the atmospheres in a given bin of column mass. For \muram we show the distribution of the temperature as a 2D histogram which is normalized to the peak of the distribution in a given bin of logarithmic column mass. For both simulations we converted the column mass scale to a geometrical height that is shown on the top horizontal axis of the plot. Panel (b): Average non-thermal velocity as function of column mass. The non-thermal velocity is computed as the standard deviation of the vertical velocity within a range of $\pm 1 \; \mathrm{dex}$ in logarithmic column mass and multiplied by a factor of $\sqrt{2}$.}
         \label{fig:temperature-and-unth}
   \end{figure}
% ++++++++++++++++++++++++++++++++++++++++++++++++++++++++++++++++++++++++++++++++++++++++++++++++++++

We applied the same degradation procedure to the synthetic \bifrost spectra as for the \muram model (\cf Sect. \ref{sec:methods_degradation}). As can be seen in Fig. \ref{fig:av_spectra} the two modeled spectra appear to be different in the peak intensities, peak separation, separation between the \kone minima, and the intensity of the "pseudo"-continuum between the two emission lines. This suggests that the structure of the modeled atmospheres are different, from the upper photosphere all the way to the transition region.

In Fig. \ref{fig:temperature-and-unth} we plot the temperature distributions of the atmospheres versus the logarithmic column mass $\log_{10} \sigma$. The colored lines represent the most likely temperature (i.e. the peak of the distribution) within a given range of logarithmic column mass. For the temperature histogram we excluded temperatures larger than $12 \, \mathrm{kK}$.

In the range $1 \geq \log_{10}\sigma\,/\, \mathrm{[kg \,m^{-2}]} \geq -1$ (approximately $1 \Mm$ above the visible surface for both models) the temperature in the \bifrost model is up to $\approx 500 \, \mathrm{K}$ higher than in the \muram model. Between $-2 \geq \log_{10}\sigma \,/\, \mathrm{[kg \,m^{-2}]} \geq -5$ the temperature in the \muram atmosphere is higher. This is the temperature range where the \mghk lines form. The distribution of the \bifrost intensities as brightness temperatures can be seen in Fig. \ref{fig:peak_separation}a. Strong NLTE effects in the formation of the peaks make a direct comparison between brightness temperature and the atmospheric temperature at the formation height difficult. \cite{Leenaarts_2013_mgii_hk_bifrost_diagnostics} showed that for the strongest peak intensities, however, there is indeed a nearly linear correlation. This suggests that the higher temperature in the \muram atmosphere at the formation heights of the line peaks lead to the higher peak intensities observed in Figures \ref{fig:av_spectra} and \ref{fig:peak_separation}a.

\cite{Carlsson2016a_public_bifrost_snapshot} proposed a way to describe the non-thermal component of the velocity field that contributes to the line broadening similar to the "microturbulence" parameter often added to 1D models to match observed line widths. The non-thermal velocity $u_{\mathrm{nth}}$ is calculated as the standard deviation of the vertical velocity in between $\pm 1 \; \mathrm{dex}$ of the logarithmic column mass $\log_{10}\sigma$ and multiplied by a factor of $\sqrt{2}$. We follow this approach and recompute $u_{\mathrm{nth}}$ for \bifrost and for our \muram spectra. The difference is small at high column mass, i.e. in the lower atmosphere, and increases toward the chromosphere and the transition region. We propose that it is mainly the larger non-thermal velocities what produces the larger line widths in \muram compared to \bifrost (see also Sect. \ref{sec:lineWidth}).

The main differences between the models are the treatment of diffusion, the horizontal spatial resolution of $24 \km$ (\muram) versus $48 \km$ (\bifrost), equidistant (\muram) versus variable (\bifrost) vertical axis, and the depth of the lower boundary of the simulation, $-7 \Mm$ (\muram) versus $-2.5 \Mm$ 
(\bifrost). The smaller diffusivity and higher resolution in \muram enable a more turbulent convection zone, leading to stronger magnetic fields. The average unsigned magnetic flux at the \tauUnity layer 
in \muram is about $64 \G$\footnote{We note that these values are much higher than the ones earlier cited. This is because these values are taken at \tauUnity and at full resolution, \ie, not degraded to the HMI resolution.} versus $48 \G$ in \bifrost.

This difference in magnetic flux likely accounts for a significant fraction of the difference in intensity between the profiles, but likely not for the difference in peak separation and line width \citep[\eg][Fig. 4]{Kayshap_2018}. We note, however, that in the synthetic spectra presented in this work, the magnetic fields do not have a direct impact on the emergent line because the Zeeman  effect is not taken into account.

In the \muram atmosphere, the lower viscosity allows for more turbulent flows and stronger shocks. The enhanced dynamics leads to a  chromosphere that extends to higher altitudes. A higher density at the formation height of the \ktwo peaks leads to a stronger coupling between the source function and the Planck function. This might explain the higher peak intensities in the \muram model.

 \cite{hansteen_numerical_mghk_2023ApJ...944..131H} report \mgk profiles from a low resolution ($100 \km$ horizontally) simulation including flux emergence which show strong broadening. The authors explain this by possible additional mass loading in the atmosphere. This supports our finding in the comparison between the \muram and \bifrost made in Fig. \ref{fig:temperature-and-unth}a where the column mass in the \muram chromosphere is roughly an order of magnitude higher than in the public \bifrost snapshot.

%====================================================================================================
\section{Summary and conclusions} \label{sec:discussion}
%====================================================================================================

In this work we modeled the \mghk lines in a quiet sun region with network elements by utilizing the chromospheric extension of the MURaM code (\muram). The simulation setup is similar to the public \bifrost snapshot and includes a bipolar magnetic field structure that was added into the simulation box after it had reached a statistically steady state. 
Although the two simulation outputs are globally similar, there are clear differences in the details. For example, structures close to the photosphere are visible on much smaller scales in the \muram simulation than in the public Bifrost snapshot.

We used a peak-finding algorithm to identify spectral features of the \mgk line such as \ktwo and \kthree. We found two qualitative discrepancies compared to the observations: 
the intensity of the  emission at the \ktwo peak is too bright above the network fields; and the shock fronts are too clear in the intranetwork regions.
Both these effects might be a shortcoming of the 1.5D RT treatment \citep{sukhorukov2017}.

We compared the spatially averaged spectra between one snapshot of our numerical model and a selected IRIS scan.  Prior to comparing the synthetic spectra with observations, we carried out a number of preparatory steps. Firstly we compared the magnetic field strength at the photosphere of the simulation to HMI measurements in the region covered by the selected IRIS scan. To this end, we first degraded the magnetic field at $\tau_{500} = 0.1$ to the HMI spatial resolution. This is approximately the height where the core of the \ion{Fe}{i} $\lambda 6173 \, \AA$ line, observed by the HMI instrument,  forms. The obtained value of $\langle | B_{z} | \rangle|_{\tau_{500}=0.1} = 20.65 \G$ is larger than in the HMI magnetogram of the observation that we used for comparison, which has a value of $\langle |B_{\mathrm{LOS} / }\mu| \rangle=12.12 \G$.

In order to compare with the IRIS data, the modeled spectra were spatially and spectrally degraded to match the resolution of the \iris observations. We find that the spectra show a reasonably good match in the separation between the \kone features and the overall line width. The peak intensities are on average higher than in the observation which could be due to the larger apparent unsigned mean magnetic field strength in the simulation. As discussed above, the possibly overestimated intensities above the network regions as a result of the 1.5D RT treatment lead also to higher peak intensities in the computed average spectrum. The averaged \muram spectra show a slightly larger blue asymmetry in both the \mgk and \mgh line than the observed spectra. This asymmetry is caused by the dominant area coverage by columns in the atmosphere that have downflows. We speculate that better results might be achieved with a larger simulation domain, which should lead to a more realistic vertical magnetic field topology. This could be achieved, for example, by extending the box size three times in each horizontal direction with the bipole being in the center of the $3 \times 3$ grid. By doing so, the bipolar structure would be more isolated from "neighboring" bipoles as a result of the horizontal periodic boundary conditions. However, the overall small but positive peak asymmetry can be seen even in smaller ROIs of the observation discussed in the main text and also in two additional observations as described in Appendix \ref{sec:app-additional-observations} and appears to be a common phenomenon.

In addition to the average spectrum we compared the distributions of \ktwo peak intensities and peak separation between the model and the observations. The distribution of the peak brightness temperatures covers the whole range of observed values. While we showed in the main text the results for the whole observed FOV, which is larger than the simulation domain, we discussed in the appendix smaller ROIs and two more data sets. There we showed that, unless the ROI contains very strong magnetic field concentrations, the distributions of the observed peak brightness temperatures are similar. The high peak brightness temperatures in the MURaM-ChE model might be explained by the different average magnetic LOS magnetic field strengths compared with the observations. Another, likely more important, contribution might come from the fact that the 1.5D RT approximation tends to overestimate peak brightness temperatures of $T_{\mathrm{b}} > 5 \, \mathrm{kK}$ and underestimate $T_{\mathrm{b}} < 5 \, \mathrm{kK}$ \citep{sukhorukov2017}. These effects might result in a more compact brightness temperature distribution which compares better with the observation.
From \citet[][Fig. 10]{sukhorukov2017} we estimate a conversion factor between the 3D RT and 1D RT average spectra of ~7.5\% in the blue peak and 12.6\% in the red peak. This, however, indicates that such a factor must be wavelength dependent. \citet[][Fig. 9]{sukhorukov2017} additionally indicate that the conversion factor should be intensity dependent. If we apply the estimated conversion factors, the peak intensities in Fig. \ref{fig:av_spectra} are still approximately 20--25\% larger than observed. A direct conversion factor from 1D RT to 3D RT spectra is, however, not straightforward to estimate.

The average peak separation of $\langle\Delta v_{\mathrm{k2}}\rangle = 23.6 \kms $ (\muram) is smaller than $\langle\Delta v_{\mathrm{k2}}\rangle = 33.02  \kms $ (\iris). Even though the observed average peak separation cannot be reached by the synthetic spectra, there is a significant overlap between the observed and simulated distributions. The  peak separation is a consequence of strong variations in the LOS velocity that arise through shocks in the atmosphere. In addition, a significant amount of non-thermal velocity in the atmosphere acts similarly as microturbulence to broaden the spectral lines.
\muram uses a slope limited numerical diffusion scheme \citep{rempel_2014_numerical,rempel_coronal_extension_2017ApJ...834...10R} that allows the simulation to run stably with a low viscosity and resistivity which leads to the observed strong velocities.

The missing peak separation might be explained by insufficient dynamics such as the difference between the minimum and maximum vertical velocity in the formation region of \mgk $\mathrm{max}(\Delta v_{z })$ in the atmosphere. This is supported by a correlation that we found \citep[similar to][]{Leenaarts_2013_mgii_hk_bifrost_diagnostics} between $\mathrm{max}(\Delta v_{z })$ and peak separations (Fig. \ref{fig:peak-separation-vgradient}), however with a slightly lower correlation coefficient, probably because of misidentifications of the peak-finding algorithm. This suggests that although the chromosphere computed by \muram is already very dynamic, the real Sun is even more dynamic. Similar to the peak asymmetry a larger simulation domain with less vertically orientated magnetic field lines, as a result of the relatively small extent of the simulation domain in the vertical direction, might lead to more realistic dynamics and larger peak separation.

We further investigated the role of velocities in their contribution to the line width. For this purpose we computed the \mghk lines for snapshot \snapfourNineNine again, but without a velocity field (\ie the velocities are set manually to zero). A similar approach was followed by, for example, \cite{Rathore_2015_opacity_broadening} who studied the effect of opacity broadening in the \ion{C}{ii} $\lambda  133.5 \nm$ line. 
By comparing single spectra from the two computations we could show that in both cases there are examples of broad lines. The increased line width even in the absence of velocities is a consequence of local maxima in the temperature in the lower atmosphere. These might be related to shocks which increase the local temperature via adiabatic heating. In the majority of the spectra we found that the presence of a velocity significantly increases the line width of the resulting spectra compared to the case without velocity. This comparison confirms the finding that chromospheric velocities play a dominant role in the line broadening of \mghk.

We tested the time-dependence of the synthesised
spectrum following two approaches. We checked for changes in 4 full
cubes saved at 2 min cadence, and in addition considered the dynamics along 8 vertical cuts (slits) sampled at 1.2 sec cadence.
From the degraded profiles we measured the peak intensities and
peak separations of the \ktwo features. The peak separations here
vary between $\langle\Delta v_{\mathrm{k2}}\rangle = 23.1 \kms $ and $\langle\Delta v_{\mathrm{k2}}\rangle = 23.6 \kms $.
The difference between the largest and smallest mean peak separation values of $0.5 \kms$ is smaller than the spectral resolution of IRIS of $\delta v = 2.7 \kms$.

The spectrum calculated from the spatial average of the 8 slits shows periodic variation in time. The variations are strongest in the peaks of the \mgk and \mgh line. This might be explained by heating of chromospheric plasma which locally increased the peak intensity according to the temperature to peak intensity correlation found by \cite{Leenaarts_2013_mgii_hk_bifrost_diagnostics}. The time series is too short to determine whether the varying intensity is a result of wave propagation in the modeled atmosphere or is due to a box mode as found in the public \bifrost snapshot \citep[see][Fig. 8]{Carlsson2016a_public_bifrost_snapshot}. 
We also found that the slit-averaged spectra further averaged over the $10 \min$ time series is rather similar to the spectrum averaged over the whole horizontal domain arising from a single snapshot. Because of the low variation in both the high-cadence slits, and the low-cadence full cubes, we conclude that the average properties of the spectrum are not dependent on the chosen snapshot. 

In summary, we find a close match between the modeled \mghk lines and spectra from IRIS observations. The synthesized spectra, however, show on average a $\approx 10 \kms$ smaller peak separation and overall slightly smaller line width. The distribution of the modeled peak intensities exceeds the observed distribution at the higher and lower intensity end. It has to be shown for this model in a future study by how much this effect is attributed to the 1.5D RT approach. Based on our results, using the 1.5D RT approach, the \muram model signifies a step forward in terms of reproducing the physical properties of the chromosphere.

 \begin{acknowledgements} 
We thank the anonymous referee for valuable comments and suggestions that improved the quality of this paper. P.O. would like to thank T. Pereira for discussions about spatial and spectral degradation to instrumental conditions and support for the HELITA\footnote{\url{https://helita.readthedocs.io/en/latest/index.html}} package, as well as the Bifrost group for discussions about the interpretation of synthetic spectral lines. P.O. would also like to thank L.P. Chitta, N. Milanovic and H. Peter for discussions about IRIS observations and alignment with HMI magnetograms. In addition, P.O. acknowledges funding from the international Max Planck Research School (IMPRS). This research has received financial support from the European Union’s Horizon 2020 research and innovation program under grant agreement No. 824135 (SOLARNET) and through the European Research Council (ERC) No. 101097844 (WINSUN). This work was supported by the Deutsches Zentrum f{\"u}r Luft und Raumfahrt (DLR; German Aerospace Center) by grant DLR-FKZ 50OU2201. We gratefully acknowledge the computational resources provided by the Cobra \& Raven supercomputer systems of the Max Planck Computing and Data Facility (MPCDF) in Garching, Germany.
 
 \end{acknowledgements} 

\bibliographystyle{aa}

\begin{appendix}
\section{Additional observations}
\label{sec:app-additional-observations}

\begin{figure*}
%\sidecaption
\centering
\includegraphics[width=\hsize,clip]{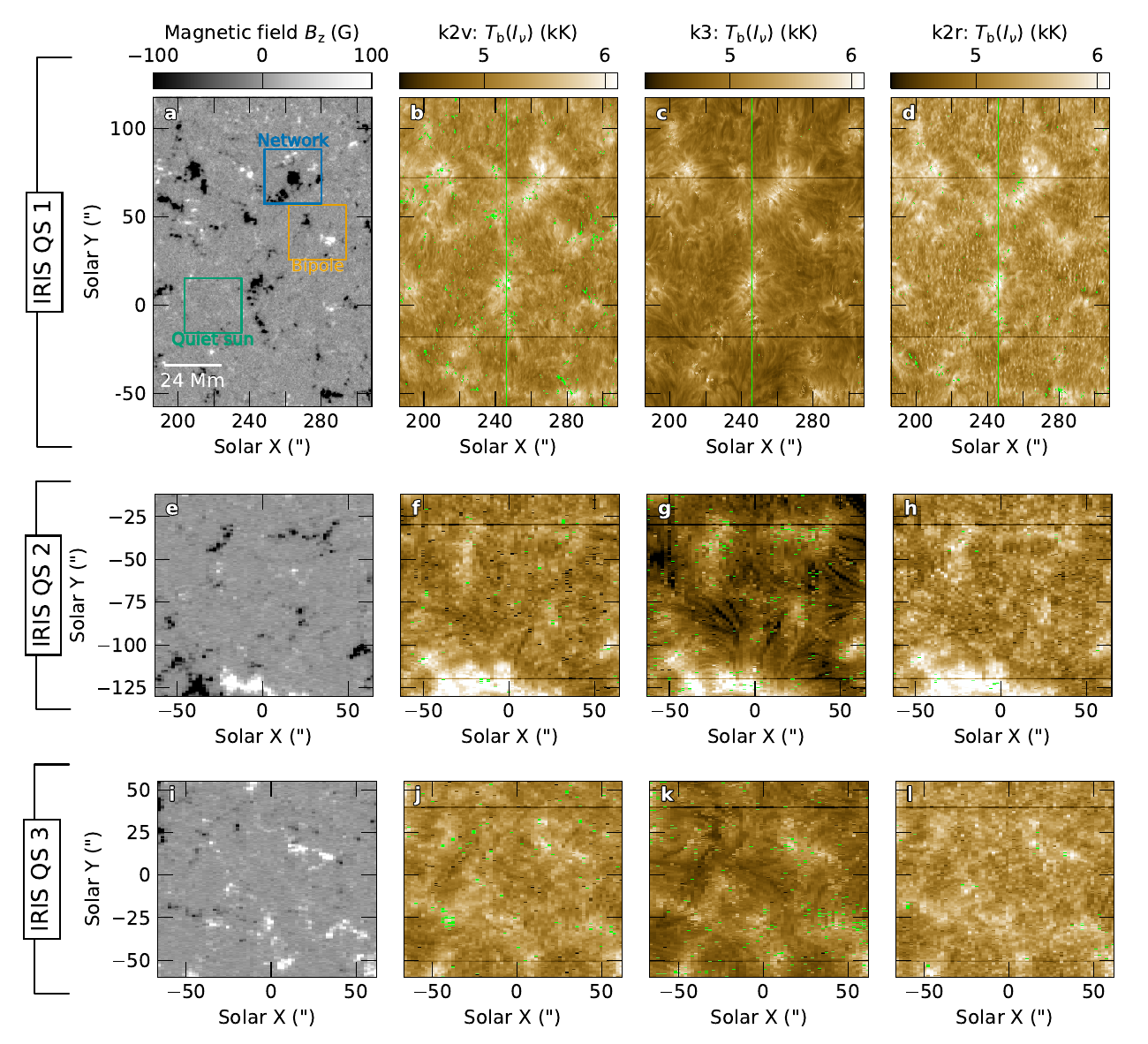}
\caption{Comparison between three IRIS datasets representing quiet sun and network regions. The first data set "IRIS QS 1" (panels a--d) is the same as the one shown in main text. Three boxes of roughly similar size as the \muram model are shown in panel (a). The second and third data set are shown in panels (e--h, IRIS QS 2) and panels (i--l, IRIS QS 3). Panels (a,e,i) show the LOS magnetic field strength measured by the HMI. Panels (b,f,j) show the brightness temperature $T_{\mathrm{b}}$ at the \ktwov, panels (c,g,k) at the \kthree and panels (d,h,l) at the \ktwor feature. Green colored pixels indicate where no feature could be detected.}
\label{fig:fig12_app_observations}
\end{figure*}

\begin{figure*}
%\sidecaption
\centering
\includegraphics[width=\hsize,clip]{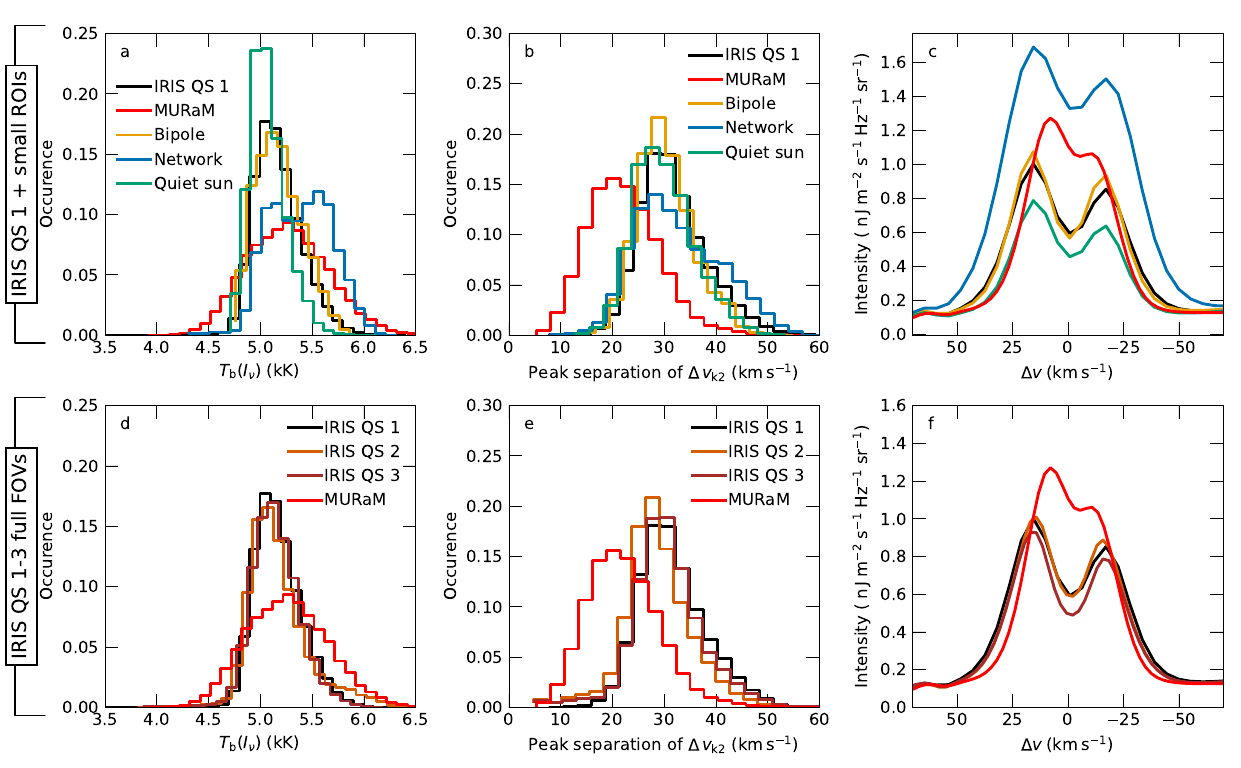}
\caption{Comparison between statistical properties of three IRIS datasets, representing quiet sun and network regions, and the \muram model. Panels (a), (b), (d), and (e) show distributions of \ktwo brightness temperature and peak separation. Panels (c) and (f) show average profiles of the \mgk line. In panels (a--c) we compare results from the first IRIS data set "IRIS QS 1" considering the full FOV together with smaller ROIs of roughly the size of the simulation domain with the results from the \muram model. In panels (d--f) we compare three IRIS data sets considering the whole FOVs together with results from the \muram model.}
\label{fig:fig13_app_observations}
\end{figure*}

\begin{table*}
\centering
\caption{Overview of observations.}
   \begin{tabular}{lcccccc}
\toprule
Description & Date & Field of view $\mathrm{(arcsec)}$ & $\mu$&$\left|B_{\mathrm{LOS}}/\mu \right|_{\mathrm{avg}}\, \mathrm{(G)}$ & $\langle T_{\mathrm{b}}\rangle \, (\mathrm{kK})$ & $\langle\Delta v_{\mathrm{k2}}\rangle \, (\mathrm{\kms})$ \\

\midrule
IRIS-QS-1 & 2014-06-07 & 139 $\times$ 182 & 0.96 & 12.12 & 5.22 & 33.02 \\
BIPOLE & 2014-06-07 & 32 $	\times$ 32 & 0.96 & 12.34 & 5.22 & 31.61 \\
NETWORK &  2014-06-07& 32 $	\times$ 32 & 0.96 & 33.20 & 5.44 & 34.06 \\
QUIET SUN &2014-06-07  & 32 $	\times$ 32 & 0.98 & 6.8 & 5.12 & 31.26 \\
IRIS-QS-2 & 2015-07-06 & 126 $\times$ 129 & 1.0 & 13.09& 5.22 & 29.96 \\
IRIS-QS-3 & 2014-05-20 & 127 $\times$ 129 & 1.0 & 8.15  & 5.20 & 31.82 \\
\bottomrule

\end{tabular}
\label{app:tab-observations}

\end{table*}
In the main text we compared the results from the \muram model with a single observation (hereafter IRIS QS 1). This observation was averaged over a FOV of $139" \times 182"$, which is larger than the simulation domain ($32" \times 32"$). We compared the average spectra in Fig. \ref{fig:av_spectra}, and the distribution of peak brightness temperatures and peak separation in Fig.  \ref{fig:peak_separation}. In this appendix we present results from smaller ROIs of $32" \times 32"$ within IRIS QS 1. These smaller ROIs compare better with the size of the model. Additionally, we present two more IRIS data sets taken on 2015-07-06 (hereafter IRIS QS 2) and 2014-05-20 (hereafter IRIS QS 3) with FOVs of $126" \times 129"$ and $127" \times 129"$ at disk center. A summary of these observations is shown in Table \ref{app:tab-observations}. In Fig. \ref{fig:fig12_app_observations} we show intensity maps and in Fig. \ref{fig:fig13_app_observations} average spectra, and the distribution of peak intensity and peak separation of the observations.

The smaller ROIs in IRIS QS 1 are indicated in Fig. \ref{fig:fig12_app_observations}a. The various ROIs show regions with differing magnetic field strengths. The first ROI contains a bipolar feature, qualitatively comparable to the magnetic configuration in our model and is therefore labeled as "bipole". The second ROI contains larger and stronger magnetic field patches and is therefore referred to as "network". The third ROI is positioned to not include any stronger magnetic field patches and therefore is meant to represent a "quiet sun", or internetwork region. In panels (a),(b), and (c) in Fig. \ref{fig:fig13_app_observations} it can be seen that the bipolar structure shows similar distributions in terms of peak brightness temperature, peak separation, and average spectrum compared with the full FOV of IRIS-QS-1. The average peak separation in the bipolar region is roughly $1.5 \kms$ smaller than in the whole FOV. The "quiet sun" region shows on average roughly $100 \, \mathrm{K}$ lower brightness temperatures than the full FOV and the average peak separation is $\approx 1.8 \kms$ smaller than in the full FOV. The "network" region contains a larger amount of stronger magnetic fields which results in a roughly three times higher averaged LOS magnetic field strength. The distribution of the corresponding peak brightness temperature, shown in panel (a), appears double peaked. The higher brightness temperature peak is a result of the bright structures that are correlated with the configuration of the photospheric magnetic field (compare Fig. \ref{fig:fig12_app_observations} panel a with panels b--d). The peak separation is around $1 \kms$ larger than in the full FOV. The overall line width of the average profiles is largest in the network region and smallest in the quiet sun region (Fig. \ref{fig:fig13_app_observations}c), while the brightness temperature scales with the LOS magnetic field strength.

The second dataset "IRIS QS 2" is presented in Fig. \ref{fig:fig12_app_observations} panels (e--h). At the lower left edge of the FOV, there is a bipolar magnetic structure which extends roughly from $(x,y) = (-50",120")$ to $(x,y) = (-50",0")$. This is larger than the size of our simulation and, additionally, it is part of a larger magnetic structure that is not part of the FOV of the raster scan. In the corresponding intensity maps, at the selected \mgk features, it can clearly be seen that there is a strong enhancement in intensity above this region.

The third dataset "IRIS QS 3" contains similarly to the other two datasets network magnetic field elements. In panels Fig. \ref{fig:fig13_app_observations}d--f we compare the statistical distributions of the three datasets taken over the total FOVs. It can be seen that the distributions of peak brightness temperatures look similar. There is, however, a small enhancement in the tail toward the higher brightness temperatures in panel (d) in IRIS QS 2, which is likely because of the above-mentioned strong bipolar field region. The average peak separation is the largest for IRIS QS 1 ($33.03 \kms$) followed by IRIS QS 3 ($31.82 \kms$) and the smallest for IRIS QS 2 ($29.96 \kms$). The overall shape of the average spectra is roughly similar between the three observations (panel f), but the intensity of IRIS QS 3 is slightly smaller than for the other two datasets.

We conclude from this study that the results from the main text, where we compared the results from \muram model with the full FOV of IRIS QS 1, are applicable to smaller ROIs and other datasets. The contrast of the peak brightness temperature in the model is slightly too high compared with the observations. The peak of the synthesized \ktwo brightness temperature distribution is between the quiet sun, with lower $B_{\mathrm{LOS}}/\mu$ values than the simulation, and observations of network elements, with higher $B_{\mathrm{LOS}}/\mu$ values than the simulation. The peak separation from the \muram model is on average roughly $10 \kms$ smaller than in the observations.  We find a similarly good match between the overall line width in all cases except for the selected network region, which has a significantly higher average longitudinal magnetic field. While the distribution of peak brightness temperatures shows a higher occurrence of brighter structures, the overall line width of the \muram model is smaller.

\end{appendix}

\end{document}